\documentclass{aa}  

\usepackage{graphicx}
\usepackage{txfonts}
\begin{document}

   \title{Contribution of phase-mixing of Alfv\'en waves to coronal\\heating in multi-harmonic loop oscillations}

   \author{P. Pagano\inst{1} \and
          D.~J.~Pascoe\inst{2,3} \and
          I. De Moortel\inst{1}
          }

\authorrunning{Pagano et al.}
\titlerunning{Heating by multi-harmonic oscillations}

   \institute{School of Mathematics and Statistics, University of St Andrews, North Haugh, St Andrews, Fife, Scotland KY16 9SS, UK\\
      \email{pp25@st-andrews.ac.uk}
   \and
      Centre for Fusion, Space and Astrophysics, Department of Physics, University of Warwick, CV4 7AL, UK
   \and
Centre for mathematical Plasma Astrophysics, Mathematics Department, KU Leuven, Celestijnenlaan 200B bus 2400, B-3001 Leuven, Belgium\\
      \email{david.pascoe@kuleuven.be}
}

   \date{ }

% \abstract{}{}{}{}{} 
% 5 {} token are mandatory

  \abstract
   {Kink oscillations of a coronal loop are observed and studied in detail because they provide a unique probe into the structure of coronal loops through MHD seismology and a potential test of coronal heating through the phase-mixing of Alfv\'en waves.
   In particular, recent observations show that standing oscillations of loops often involve also higher harmonics, beside the fundamental mode. The damping of these kink oscillations is explained by mode coupling with Alfv\'en waves.}
   {We investigate the consequences for wave-based coronal heating of higher harmonics and what coronal heating observational signatures we may use to infer the presence of higher harmonic kink oscillations.}
   {We perform a set of non-ideal MHD simulations where the damping of the kink oscillation of a flux tube via mode coupling is modelled. Our MHD simulation parameters are based on the seismological inversion of an observation for which the first three harmonics are detected.
   We study the phase-mixing of Alfv\'en waves that leads to the deposition of heat in the system, and we apply the seismological inversion techniques to the MHD simulation output.
   }
   {We find that the heating due to phase-mixing of the Alfv\'en waves triggered by the damping of the kink oscillation is relatively small,
   however we can illustrate i) how the heating location drifts due to
   the subsequent damping of lower order harmonics.
   We also address the role of the higher order harmonics and the width 
   of the boundary shell in the energy deposition.}
   {We conclude that the coronal heating due to phase-mixing seems
   not to provide enough energy to maintain the thermal structure of the solar corona even when multi-harmonics oscillations are included,
   and these oscillations play an inhibiting role in the development
   of smaller scale structures.}

   \keywords{Magnetohydrodynamics (MHD) -- Sun: atmosphere -- Sun: corona -- Sun: magnetic fields -- Sun: oscillations -- Waves}

   \maketitle
%
%________________________________________________________________

\section{Introduction}
\label{sect:intro}

%\begin{itemize}
%\item{Coronal Heating problem}
%\item{Loops dynamics}
%\item{Role of waves and standing waves in coronal heating}
%\item{KHI}
%\item{Observation of more harmonics}
%\item{Need to put the harmonics in the picture to understand how they can contribute}
%\end{itemize}

The solar corona is a highly dynamic and complex environment where magnetic fields play a key role in shaping coronal structures.
Specifically, magnetic loops are ubiquitous in the solar corona and are present in different lengths and sizes, even if their internal structure remains elusive. \citet{Reale2010} provides a review on the nature of these structures.
At the same time, owing to their relatively simple structure, they represent a unique laboratory to test coronal heating models. To this end coronal seismology is a great support for these investigations as it provides essential information about the properties and geometry of these structures.
\citet{DeMoortelBrowning2015} and \citet{ParnellDeMoortel2012}
provide an overview of the coronal heating problem and the open questions that still need to be addressed.

Many models of coronal heating rely on the conversion of Alfv\'en wave energy into thermal energy
\citep[see review by][]{Arregui2015},
and in particular the roles of phase-mixing of Alfv\'en waves \citep{HeyvaertsPriest1983}
and resonant absorption \citep{Goossens1992,Goossens2002}
have been scrutinised.

For these reasons, oscillations of coronal loops have been studied in detail to derive information on the internal structure of the loop and the  processes ongoing during the oscillations
\citep{Aschwanden1999,Aschwanden2002,Aschwanden2011}.
Some oscillations are long lived, with minimal damping \citep[e.g.][]{Nistico2013,Anfinogentov2013,Anfinogentov2015,Antolin2016},
while in other circumstances 
observations clearly show damped oscillations
\citep[e.g.][]{Nakariakov1999,Nistico2013,2016A&A...589A.136P,Pascoe2016b}.
Numerical studies \citep[e.g.][]{2008ApJ...679.1611T,Pascoe2010,Pascoe2011} have shown that the mode-coupling is a viable mechanism to concentrate
Alfv\'en wave energy in thin boundary shells where
small scale structures can form and have suggested that the phase-mixing can then convert the wave energy.
However, recently \citet{2016A&A...586A..95T} has claimed that
phase-mixing induced-heating results significantly reduced
when taking into accounts plasma flows.

\citet{DeMoortelNakariakov2012} provide a 
review of the recent achievement of coronal seismology and \citet{Brooks2012} discuss the structure of coronal loops.
%\citep{Reale2016}
In particular, by means of the seismology inversion technique 
described in \citet{Pascoe2013} which takes into account the Gaussian regime of resonant absorption \citep{Pascoe2012,Pascoe2013,Hood2013,2013A&A...555A..27R},
it was possible to derive the properties of the loop and the background corona in which the loop oscillation takes place \citep{2016A&A...589A.136P}.

Additionally, there have been indications
that the triggering of loop oscillation does not only involve 
the fundamental kink oscillation mode, but
higher parallel harmonics are possibly excited
\citep{DeMoortelBrady2007,Wang2008,VanDoorsselaere2007b,2016A&A...593A..53P}.
\citet{2016A&A...589A.136P,2017A&A...600A..78P} performed detailed analysis of loop oscillations from the catalogue of \citet{ZimovetsNakariakov2015,Goddard2016}.
The oscillations of the loop were observed with the instrument AIA on board SDO \citep{Lemen2012} and triggered by a nearby eruption occurring.
Using Bayesian analysis, \cite{2017A&A...600A..78P} discovered evidence of the second and/or third parallel harmonics being excited for kink oscillations generated by external perturbations \citep[consistent with the numerical simulations by][]{2009A&A...505..319P,2014ApJ...784..101P}. However, the same analysis found evidence against the existence of higher parallel harmonics for a kink oscillation generated by the post-flare implosion \citep{2017A&A...607A...8P}.

These studies show that multiple harmonics oscillations are not uncommon in the solar corona and are detectable with current instruments and modelling techniques.
Therefore, considering the ongoing investigation on the wave based heating mechanisms and the recent observations of multi-harmonics loop oscillations, we aim to study the effects and observational signatures of additional harmonics on coronal heating.

To carry out our investigation we devise a set of non-ideal MHD simulations in which we model a coronal loop with an inhomogeneous magnetic flux tube whose properties are based on the loop system whose observations are discussed in \citet{2017A&A...600A..78P}.
Further details on the links between the modelling and the observation can be found in Sect.~\ref{modelconobs}.
We prescribe a transverse velocity profile along the magnetic flux tube to model the excitation of standing kink oscillations, where the parameters of the initial velocity field are chosen to reproduce the observed multi-harmonic oscillations.
The resulting kink oscillations are composed of the first three harmonics of the magnetic flux tube system and they undergo mode coupling by resonant absorption, leading to phase mixing and plasma heating.
In order to highlight the role of the higher order harmonics in coronal heating, we consider the heating distribution in time and space correlated with the evolution of each harmonic. From this analysis we derive potential observational signatures of multi-harmonic oscillations related to coronal heating, also comparing different MHD simulations with a different number of harmonics.

The paper is structured as follows. In Sect.~\ref{modelconobs} we address our MHD modelling and the observation on which particular loop parameters are based. In Sect.~\ref{referencesim} we describe the MHD simulations we carry out. In Sect.~\ref{compobs} we carry out a seismology inversion on the output of the MHD simulation and we compare the results with the observations, and we finally discuss results and draw some conclusions in Sect.~\ref{conclusion}.

\section{Model and connection with observation}
\label{modelconobs}

In order to describe the transverse kink oscillation of a coronal loop and the subsequent mode coupling with Alfv\'en waves and phase-mixing, we devise a model with a magnetised cylinder anchored at each end.
The density and the Alfv\'en speed vary across the cylinder and the initial velocity field is used to trigger the transverse kink oscillations.
The properties of the flux tube and its oscillation
are based on "Loop $\#1$" which was first analysed seismologically in \citet{2016A&A...589A.136P}, and later in \citet{2017A&A...600A..78P} which extended the method, in particular by considering the presence of parallel harmonics in addition to the fundamental standing mode.
Details follow in Sect.~\ref{sect:initial} and Sect.~\ref{sect:kinkosc} for the initial flux tube properties and the initial kink oscillations, respectively.

\subsection{Initial Condition}
\label{sect:initial}

\begin{figure}
\centering
\includegraphics[scale=0.5,clip,viewport=80 68 370 645]{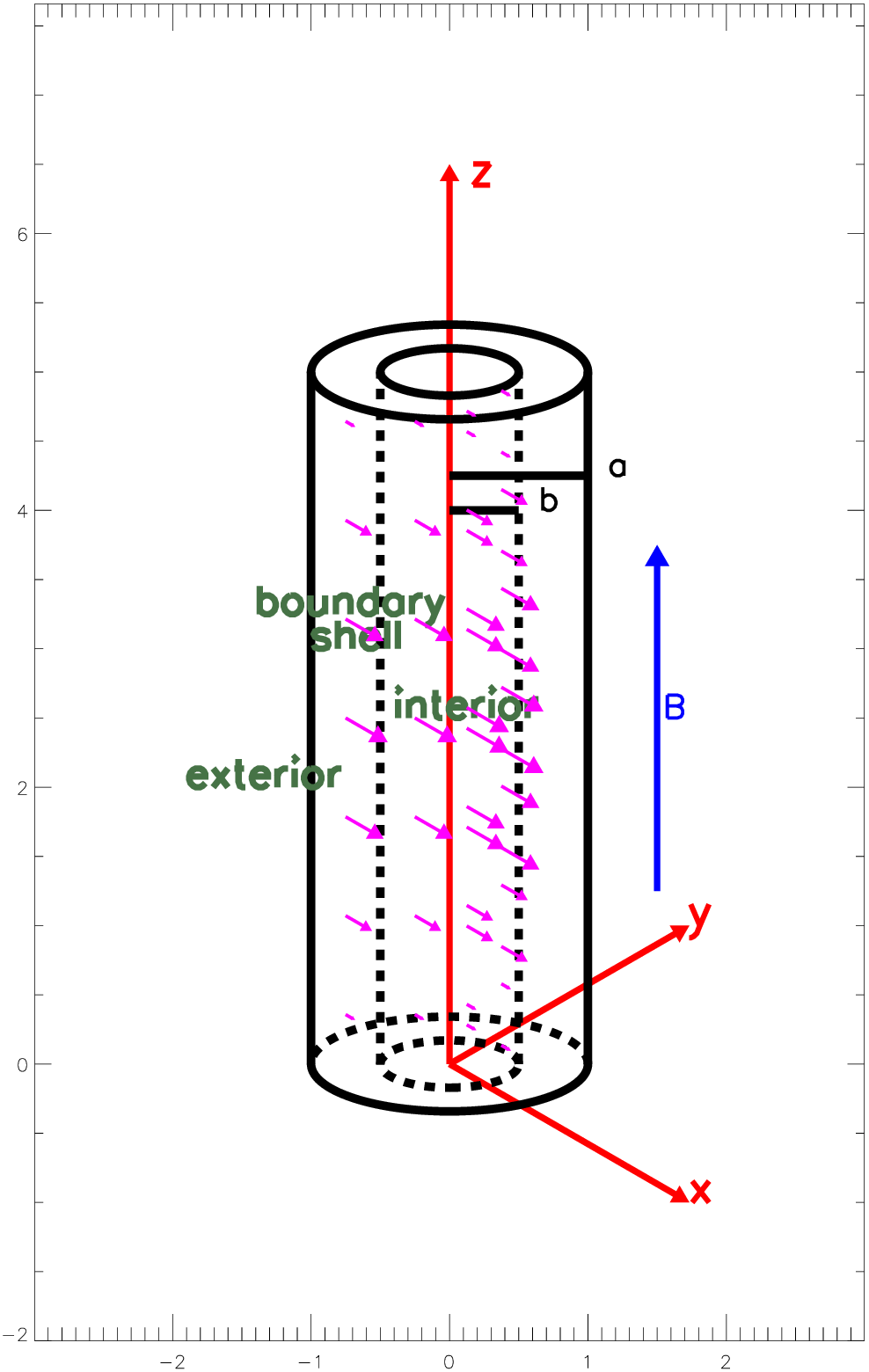}
\caption{Sketch to illustrate the geometry of our system and the Cartesian axes (red arrows).
The blue arrow represents the direction of the magnetic field and the magenta arrows represent the plasma velocity vectors
associated with a kink oscillation. The black lines are to identity the different regions of the loop.}
\label{sketch}
\end{figure}

We consider a cylindrical flux tube where we define an interior region, a boundary shell, and an exterior region (Fig.\ref{sketch}).
The system is set in a Cartesian reference frame with $z$ being the direction along the cylinder axis, and $x$ and $y$ define the plane across the cylinder cross section.
The origin of the axes is placed at the centre of the cylinder, which corresponds to the loop apex.

\begin{table}
\caption{Parameters {obtained from the observations and used for our loop modelling.}}             % title of Table
\label{tableparameters}      % is used to refer this table in the text
\centering                          % used for centering table
\begin{tabular}{c c c}        % centered columns (4 columns)
\hline\hline                 % inserts double horizontal lines
Parameter & value & Units  \\    % table heading 
\hline                        % inserts single horizontal line
   $L$ & $220$ & $ Mm $  \\  
   $R$ & $1.5$ & $ Mm $  \\  
   $\epsilon$ & $1.15 $ & $ $  \\  
   $a=R-0.5\epsilon R$ & $0.925$ & $ Mm $ \\
   $b=R+0.5\epsilon R$ & $2.075$ & $ Mm $ \\
   $V_A$ & $ 1.8 $ & $ Mm/s $  \\  
   $\rho_c$ & $ 1.7 $ & $ $ \\  
   $T_0$ & $ 1$ & $MK$ \\
   $B_0$ & $ 10 $ & G \\

\hline                                   %inserts single line
\end{tabular}
\end{table}

In the seismological analysis performed on "Loop $\#1$" in \citet{Pascoe2016b}, the observations help to constrain the geometrical properties of the loop, such as the loop length, radius, and size of the boundary shell, the plasma and magnetic field properties,
and the density contrast between the interior and exterior region.
The loop length and minor radius were estimated using SDO/AIA $171${\AA} images, which were also used to create time-distance (TD) map showing the loop oscillation. Fitting the position of the loop produced a time series for the loop motion which allows seismological analysis through the measurement of the period of oscillation and shape of the damping profile. Based on the interpretation in terms of a kink oscillation damped by resonant absorption, these oscillation properties were used to calculate seismological estimates for the loop density contrast ratio and width of the boundary shell.
These values are summarised in Table~\ref{tableparameters}, where $L$ is the length of the cylinder, $R$ is the radius of the cylinder, and $\epsilon$ is the ratio between the width of the boundary shell and the radius.
We relate the radius of the interior region, $a$, and boundary shell region, $b$ as $a=R-0.5\epsilon R$ and $b=R+0.5\epsilon R$.
$V_A$ is the Alfv\'en speed in the exterior region, $B_0$ the magnetic field intensity in the same region,
and $\rho_c$ the density contrast ratio between the interior and exterior region.
We can derive the density in the exterior region,
$\rho_e=B_0^2/(4\pi V_A)$, and in the interior, $\rho_i=\rho_e \rho_c$.
Finally, we uniformly assign the temperature $T_0$.

In our simulations we model the boundary shell using the same density profile as in \citet{PaganoDeMoortel2017}

\begin{equation}
\label{densitylayer}
\displaystyle{\rho(\rho_e,\rho_i,a,b)=\rho_e+\left(\frac{\rho_i-\rho_e}{2}\right)\left[1-tanh\left(\frac{e}{a-b}\left[r-\frac{b+a}{2}\right]\right)\right]}.
\end{equation}
We note that the density profile in Eq.~(\ref{densitylayer}) (depending on $tanh$) is different to the linear transition layer density profile used in the analysis by \citet{2016A&A...589A.136P}, and so the definition of the width of the boundary shell.
By comparing the two different density profiles we find that the profile above is equivalent to a linear transition layer density profile that is 
$\sim10\%$ thinner than the one used in this simulation, for which the damping of kink oscillations by mode coupling will be slightly weaker.

\begin{figure}
\centering
\includegraphics[scale=0.28]{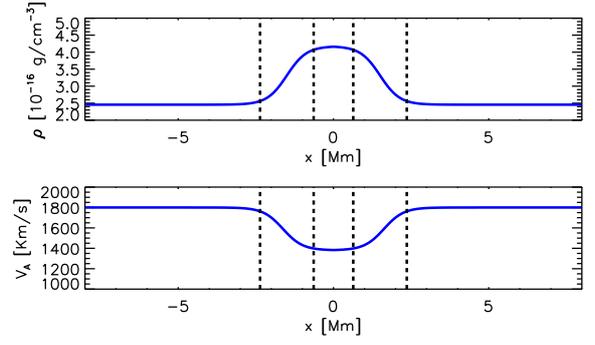} % printer
\caption{Profile of density and Alfv\'en velocity across the cylinder.
Dashed lines mark the borders of the boundary shell $a$ and $b$.
The radius of the cylinder $R$ correspondes to the centre of the boundary shell.}
\label{initdensity}
\end{figure}

The magnetic field is set to have a uniform strength $B_0$ and directed along the $z$-direction, and we set a uniform thermal pressure across the flux tube

\begin{equation}
p_e=\frac{\rho_e}{0.5 m_p} k_b T_e
\end{equation}
where $m_p$ is the proton mass and $k_b$ is the Boltzmann constant.
These assumptions lead to a
magnetohydrostatic configuration where the 
temperature varies across the flux tube.
Fig.~\ref{initdensity} shows density and Alfv\'en speed profile across the centre of the flux tube.

\subsection{Kink oscillation}
\label{sect:kinkosc}

To trigger the transverse kink oscillation in the flux tube, we impose a velocity field in the domain, that corresponds to the linear superposition of the first three standing modes along the extension of the flux tube.
The velocity field is uniform within the interior and boundary shell region, and the plasma is at rest in the exterior region.

Therefore, in our coordinate system the velocity field inside the interior and boundary shell region
only depends on the coordinate $z$ and the three standing modes ($n=1$, $2$, $3$) can be written as
\begin{equation}
f_n \left( z \right) = \sin\left(\frac{z - L/2}{\lambda_n}2\pi\right),
\end{equation}
where

\begin{equation}
\lambda_{n}=\frac{2L}{n}.
\end{equation}

In \citet{Pascoe2016b,2017A&A...600A..78P} the transverse displacement of the Loop \#1 is followed at $z \approx 0.4L$ and in this way it is possible to derive the period of the fundamental mode (the 1st harmonic) and the spatial amplitude of each standing mode.
The measured oscillation period of the fundamental mode is $P_1=280$~s for the fundamental mode,
and consequently the periods for the second ($P_2=P_1/2$) and third mode ($P_3=P_1/3$) can be calculated.
The amplitudes for the first three standing modes are estimated as $A_1=1.05$~Mm, $A_2=0.35$~Mm, and $A_3=0.15$~Mm.

\begin{figure}
\centering
\includegraphics[scale=0.28]{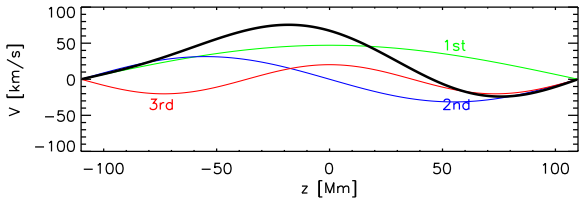} % printer
\caption{Initial velocity profile along the flux tube (black line) and the contribution from each of the three parallel harmonics:
fundamental (green), second (blue), and third (red).}
\label{initkink}
\end{figure}

In order to model the kink oscillation with an initial velocity field, we transform these spatial displacement amplitude into
velocity amplitudes by deriving the maximum displacement of the centre of the flux tube from its rest position
from the simple kinematic relation:
\begin{equation}
A_n\left(z\right)=\int_{0}^{P_n/4} V_n cos\left(\frac{t}{P_n}2\pi\right) dt = V_n \frac{P_n}{2\pi},
\end{equation}
where $t$ is time and $V_n$ is the oscillation velocity amplitude of the $n$th harmonic.
This gives:

\begin{equation}
V_n=A_n\left(z\right)\frac{2\pi}{P_n}.
\end{equation}
However, this simple derivation does not take into account
the geometry of the flux tube, the reaction of the plasma surrounding the tube, and the inevitable numerical effects due to the finite resolution.
From a test simulation we empirically conclude that we need to multiply $V_n$, by a factor of $2$ to match the modelled displacement at $z=0.4L$ with the measured value from observation.
Table~\ref{tableoscil} summarises the parameters for the initial velocity field (already multiplied by $2$).
Figure~\ref{initkink} shows the initial velocity profile along the flux tube that triggers the kink oscillation.

\begin{table}
\caption{Parameters oscillation}             % title of Table
\label{tableoscil}      % is used to refer this table in the text
\centering                          % used for centering table
\begin{tabular}{c c c c c c}        % centered columns (6 columns)
\hline\hline                 % inserts double horizontal lines
Parameter & value & Units & Parameter & value & Units \\    % table heading 
\hline                        % inserts single horizontal line
   $P_1$ & $280$ & $ s $  & $V_1$ & $47.12$ & $ km/s $ \\
   $P_2$ & $140$ & $ s $  & $V_2$ & $31.42$ & $ km/s $ \\  
   $P_3$ & $93.3$ & $ s $ & $V_3$ & $20.20$ & $ km/s $ \\    

\hline                                   %inserts single line
\end{tabular}
\end{table}

\subsection{MHD numerical setup}

In order to study the evolution of the system after the transverse kink oscillations are initiated,
we used the MPI-AMRVAC software~\citep{Porth2014} to solve the MHD equations,
where thermal conduction, magnetic diffusion, and joule heating are treated as source terms:
\begin{equation}
\label{mass}
\displaystyle{\frac{\partial\rho}{\partial t}+\vec{\nabla}\cdot(\rho\vec{v})=0},
\end{equation}
\begin{equation}
\label{momentum}
\displaystyle{\frac{\partial\rho\vec{v}}{\partial t}+\vec{\nabla}\cdot(\rho\vec{v}\vec{v})
   +\nabla p-\frac{\vec{j}\times\vec{B}}{c}=0},
\end{equation}
\begin{equation}
\label{induction}
\displaystyle{\frac{\partial\vec{B}}{\partial t}-\vec{\nabla}\times(\vec{v}\times\vec{B})=\eta\frac{c^2}{4\pi}\nabla^2\vec{B}},
\end{equation}
\begin{equation}
\label{energy}
\displaystyle{\frac{\partial e}{\partial t}+\vec{\nabla}\cdot[(e+p)\vec{v}]=-\eta j^2-\nabla\cdot\vec{F_c}}, %-n^2\chi(T),
\end{equation}
where $\vec{v}$ is the velocity,
$\eta$ the magnetic resistivity, $c$ the speed of light, $j=\frac{c}{4\pi}\nabla\times\vec{B}$ the current density, and
$F_c$ the conductive flux \citep{Spitzer1962}.
The total energy density $e$ is given by

\begin{equation}
\label{enercouple}
\displaystyle{e=\frac{p}{\gamma-1}+\frac{1}{2}\rho\vec{v}^2+\frac{\vec{B}^2}{8\pi}},
\end{equation}
where $\gamma=5/3$ denotes the ratio of specific heats.
In our numerical experiments we
adopted a value of $\eta$ that is set uniformly
as $\eta=10^9\eta_S$,
where $\eta_S$ is the classical value at $T=2$~MK \citep{Spitzer1962}.

In order to verify that this value of resistivity leads to appreciable effects above the numerical diffusivity we performed a test simulation with $\eta=0$.
We found that a value of $\eta=10^9\eta_S$ is sufficient to
develop significant effects.
To better address the evolution of the plasma during the MHD simulation we use tracers to follow the evolution of the plasma that is initially in
the interior region ($tr_i$),
the boundary shell ($tr_{bs}$),
and the exterior region ($tr_e$).
The tracers $tr_i$, $tr_{bs}$, $tr_e$ can have values between $0$ when the tracer is absent
and $1$ when only the tracer is present and $tr_i+tr_{bs}+tr_e=1$.

The computational domain is composed of $512\times320\times256$ cells, distributed on a uniform grid.
The simulation domain extends from $x=-8$~Mm to $x=8$~Mm,
from $y=-10$~Mm to $y=0$~Mm (where we model only half of a flux tube)
and from $z=-110$~Mm to $z=110$~Mm in the direction of the initial magnetic field.
The boundary conditions are implemented using ghost cells,
and we set periodic boundary conditions in $x$,
reflective boundary conditions at the $y$ boundary crossing the centre of the flux tube, and
outflow boundary conditions at the other $y$ boundary, 
and fixed boundary conditions at both $z$ boundaries
coherently with the oscillation of the standing modes.

\section{Results of MHD Simulations}
\label{referencesim}
%\begin{itemize}
%\item{General description}
%\item{Evolution of power spectrum}
%\item{Presence of heating?}
%\end{itemize}

In our study we run a MHD simulation of a transverse kink oscillation of a flux tube where the density and Alfv\'en speed vary across a boundary shell and the initial kink velocity field is the result of the contribution from the first three harmonics.
We focus our analysis on the generation of Alfv\'en waves (azimuthal oscillations) in the boundary shell and on the following phase-mixing and the dissipation of wave energy as heat.

\begin{figure}
\centering
\includegraphics[scale=0.28,clip,viewport=010 50 345 650]{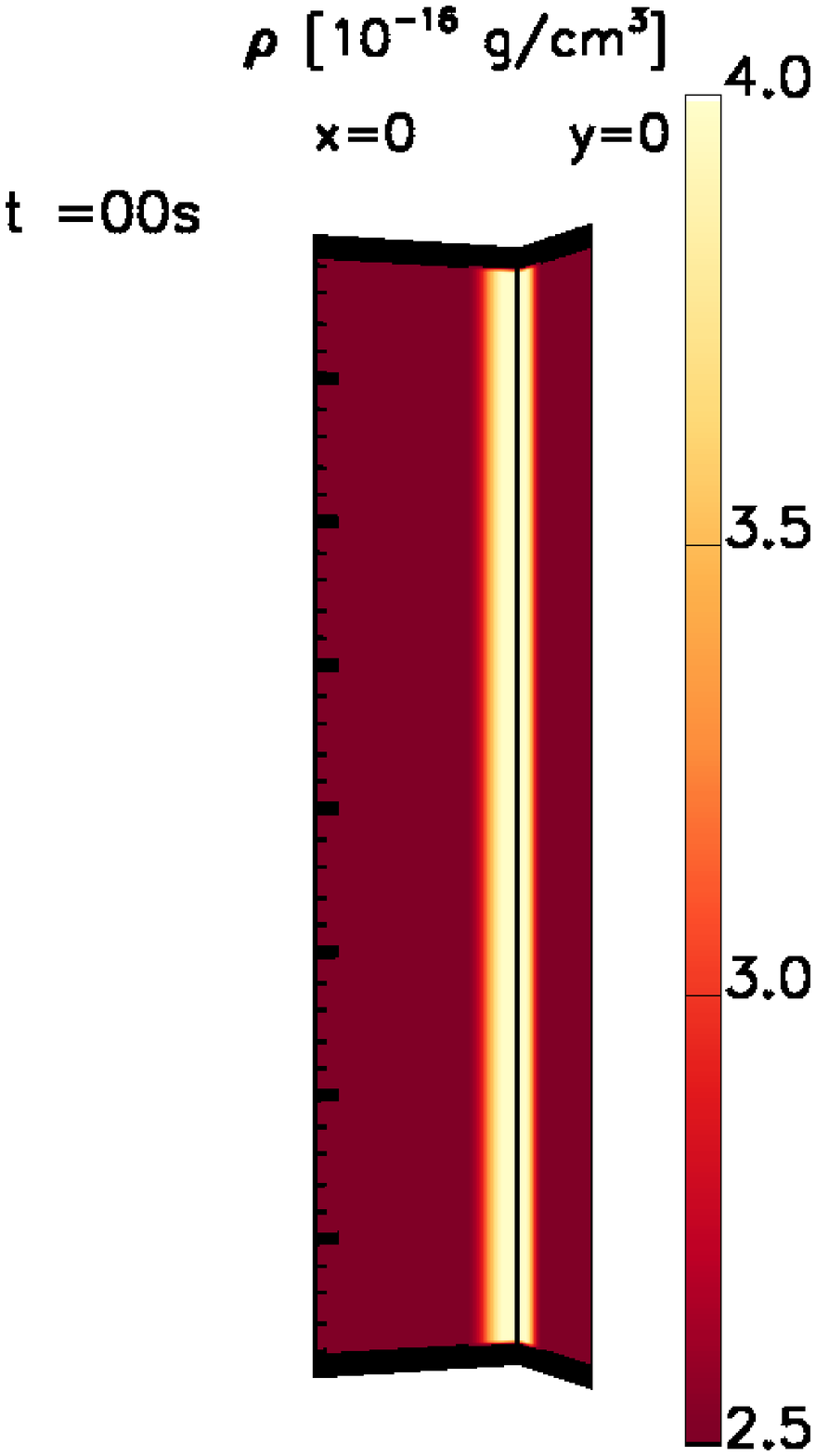} % printer
\includegraphics[scale=0.28,clip,viewport=110 50 380 650]{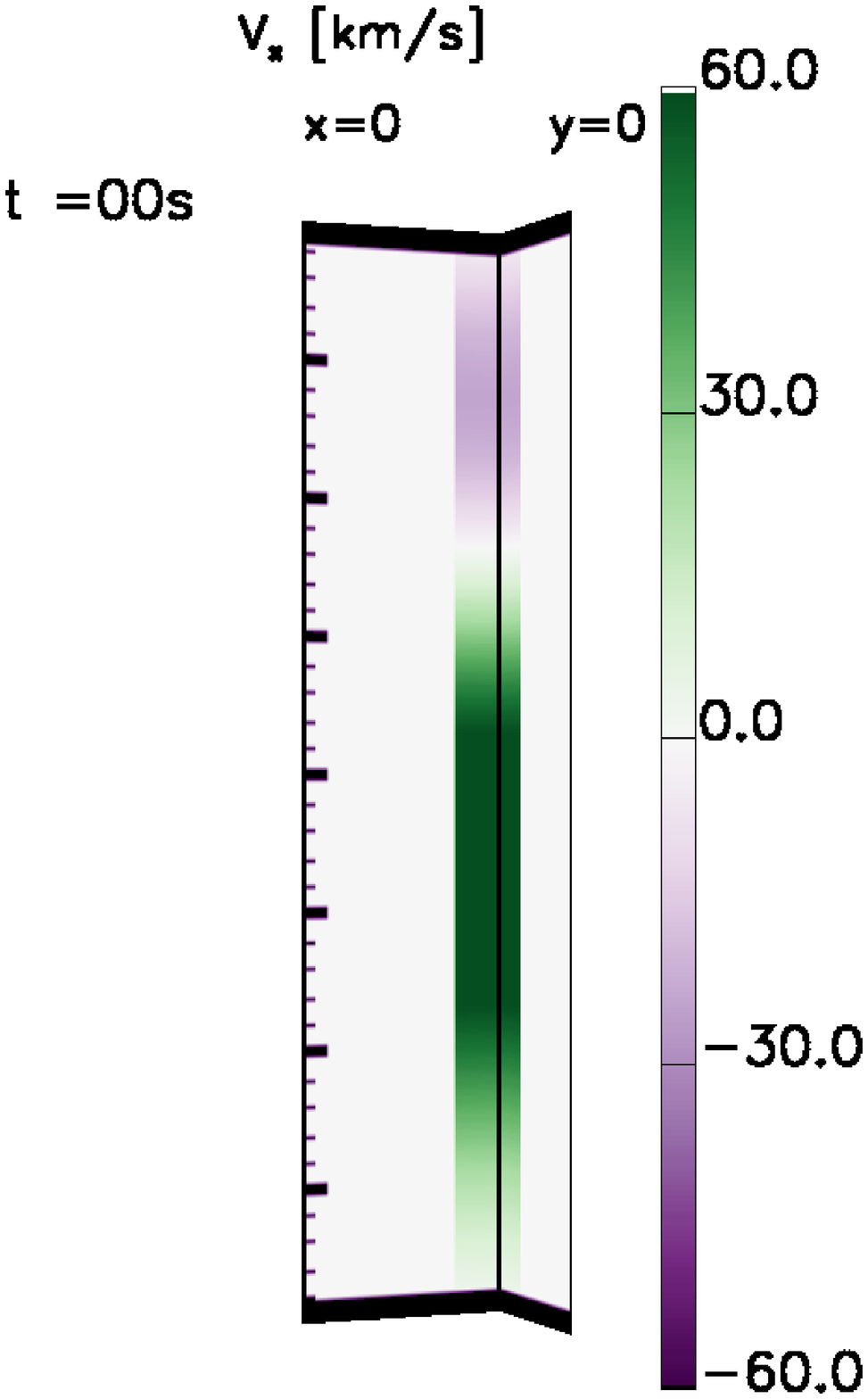} % printer
\includegraphics[scale=0.28,clip,viewport=110 50 290 650]{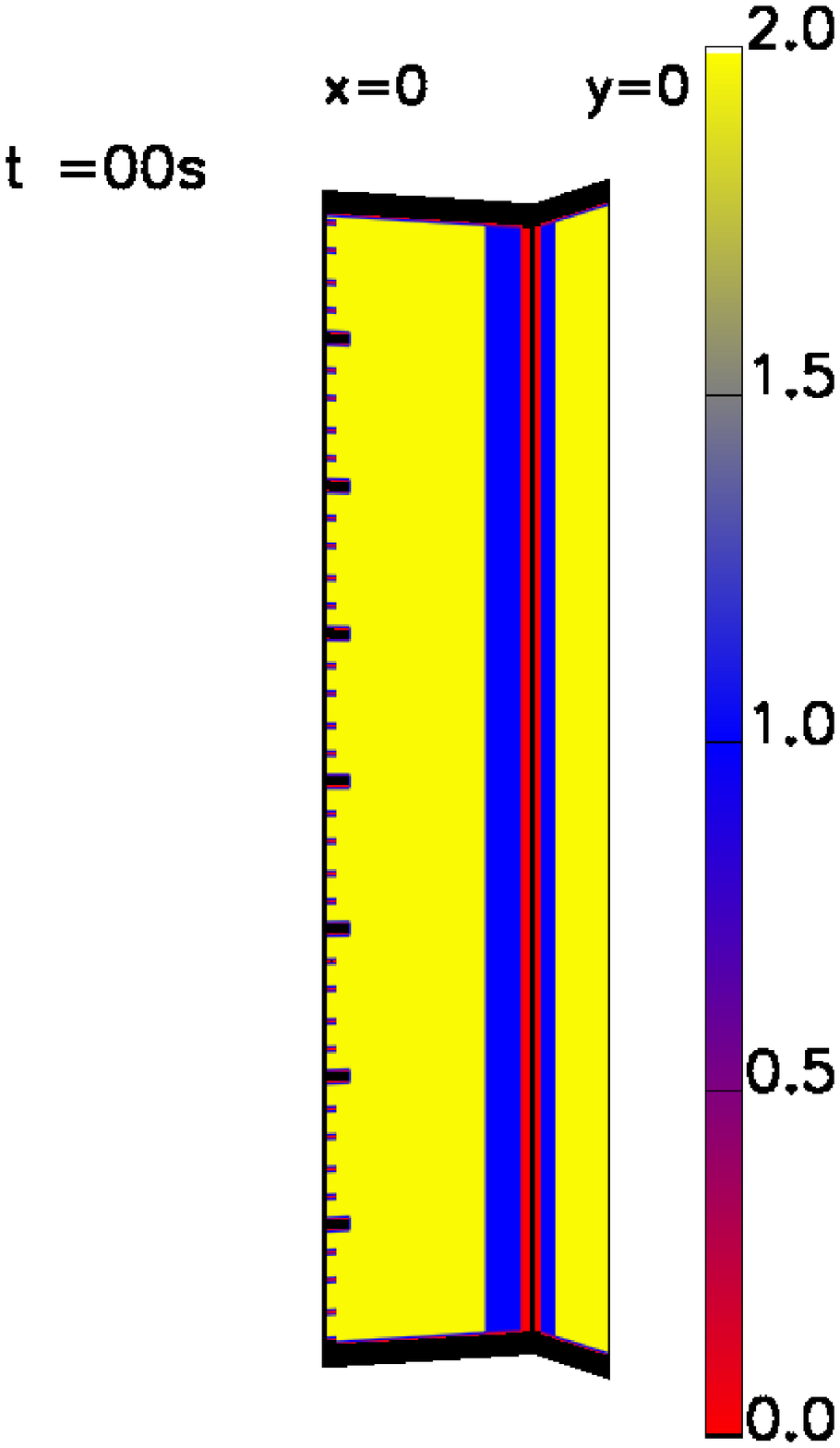} % printer

\includegraphics[scale=0.28,clip,viewport=010 50 345 650]{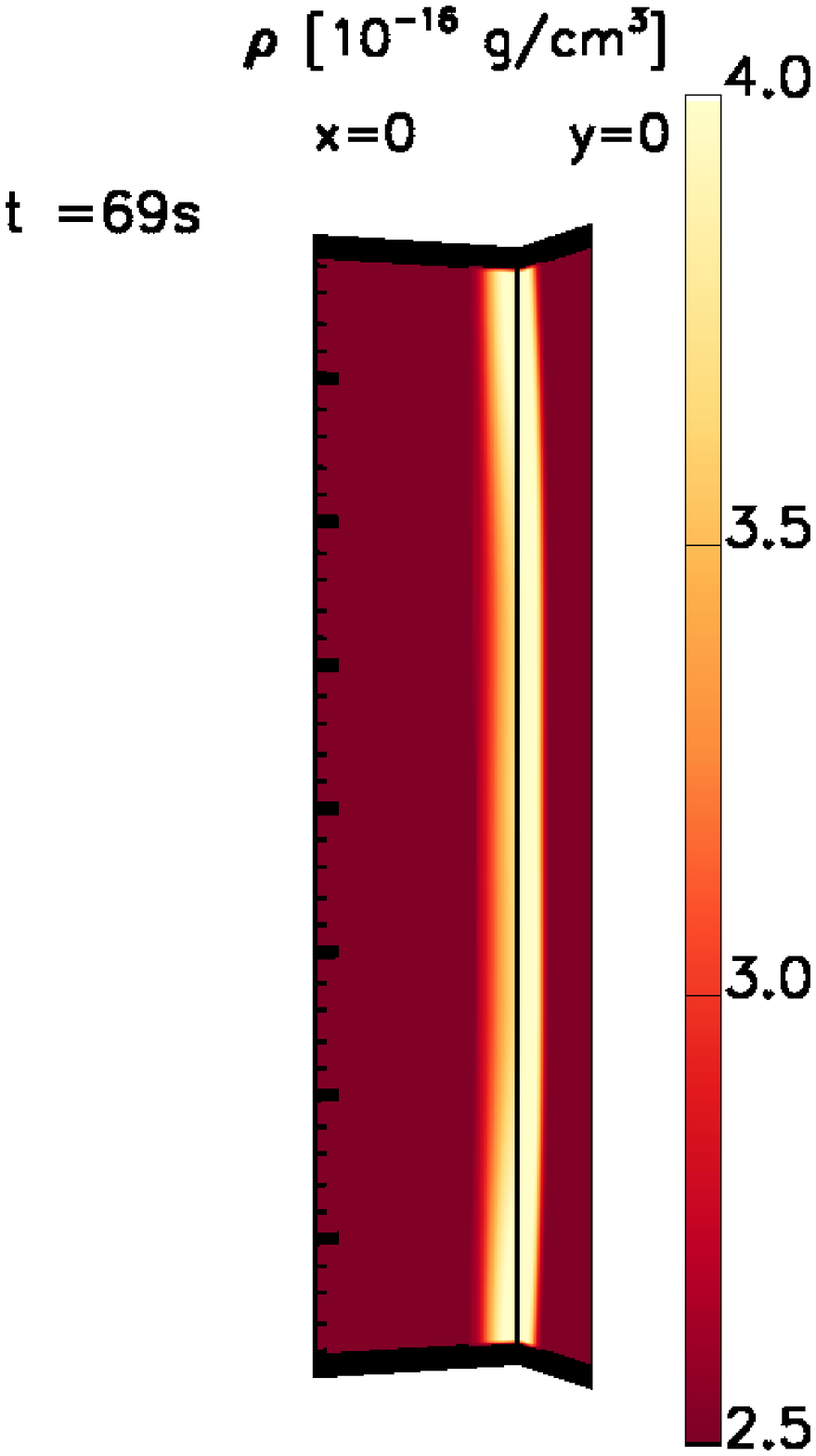} % printer
\includegraphics[scale=0.28,clip,viewport=110 50 380 650]{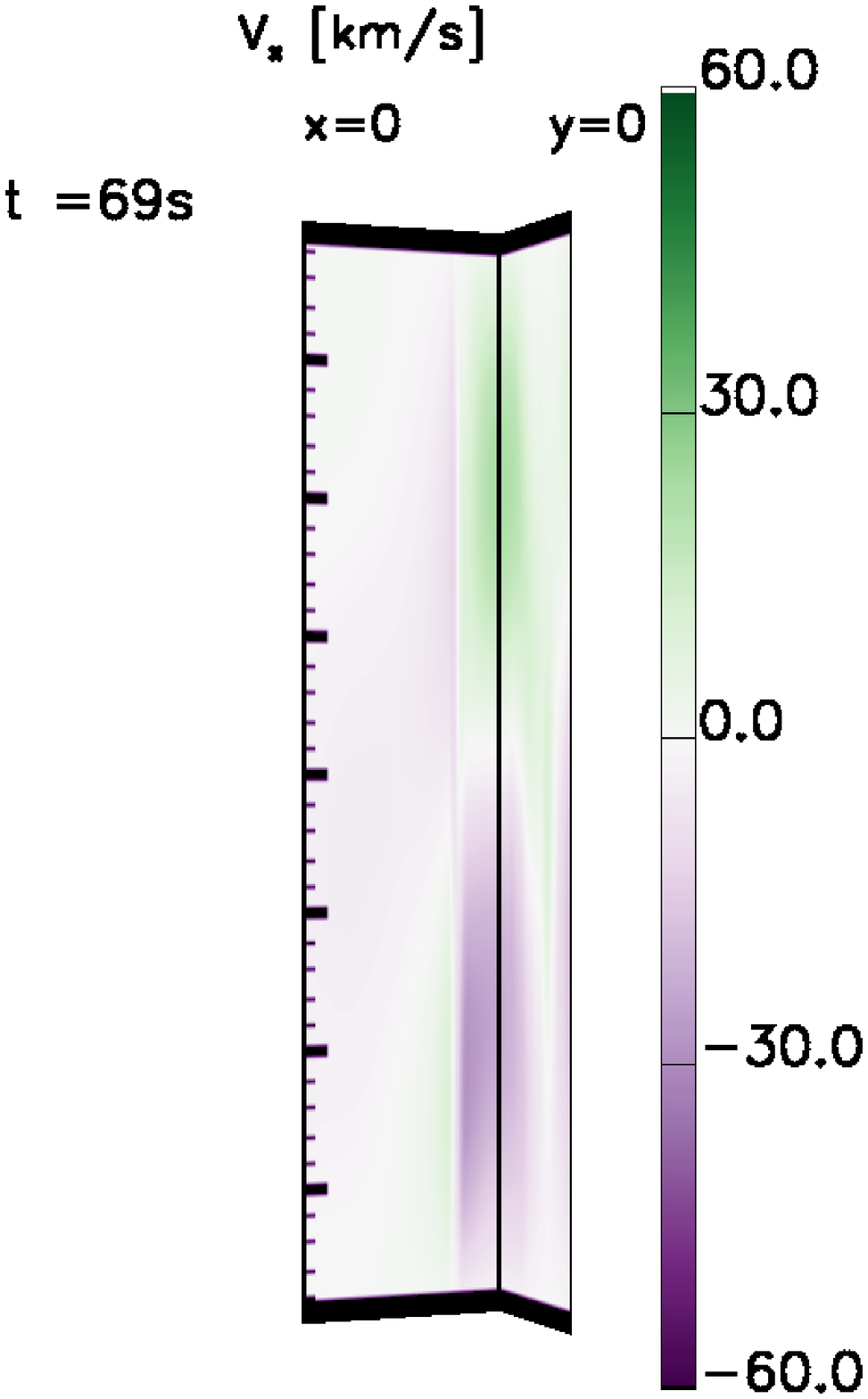} % printer
\includegraphics[scale=0.28,clip,viewport=110 50 290 650]{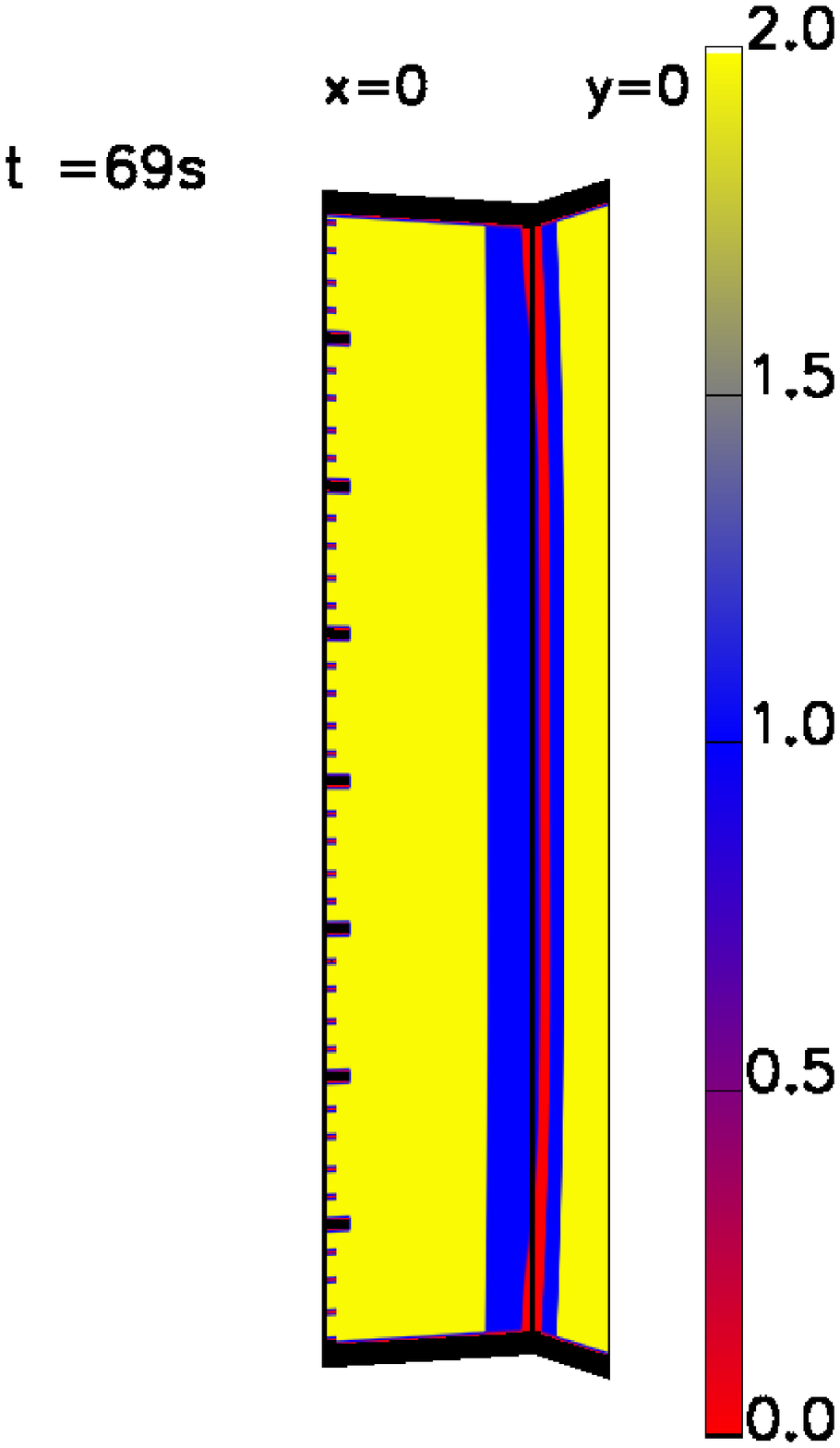} % printer

\includegraphics[scale=0.28,clip,viewport=010 50 355 650]{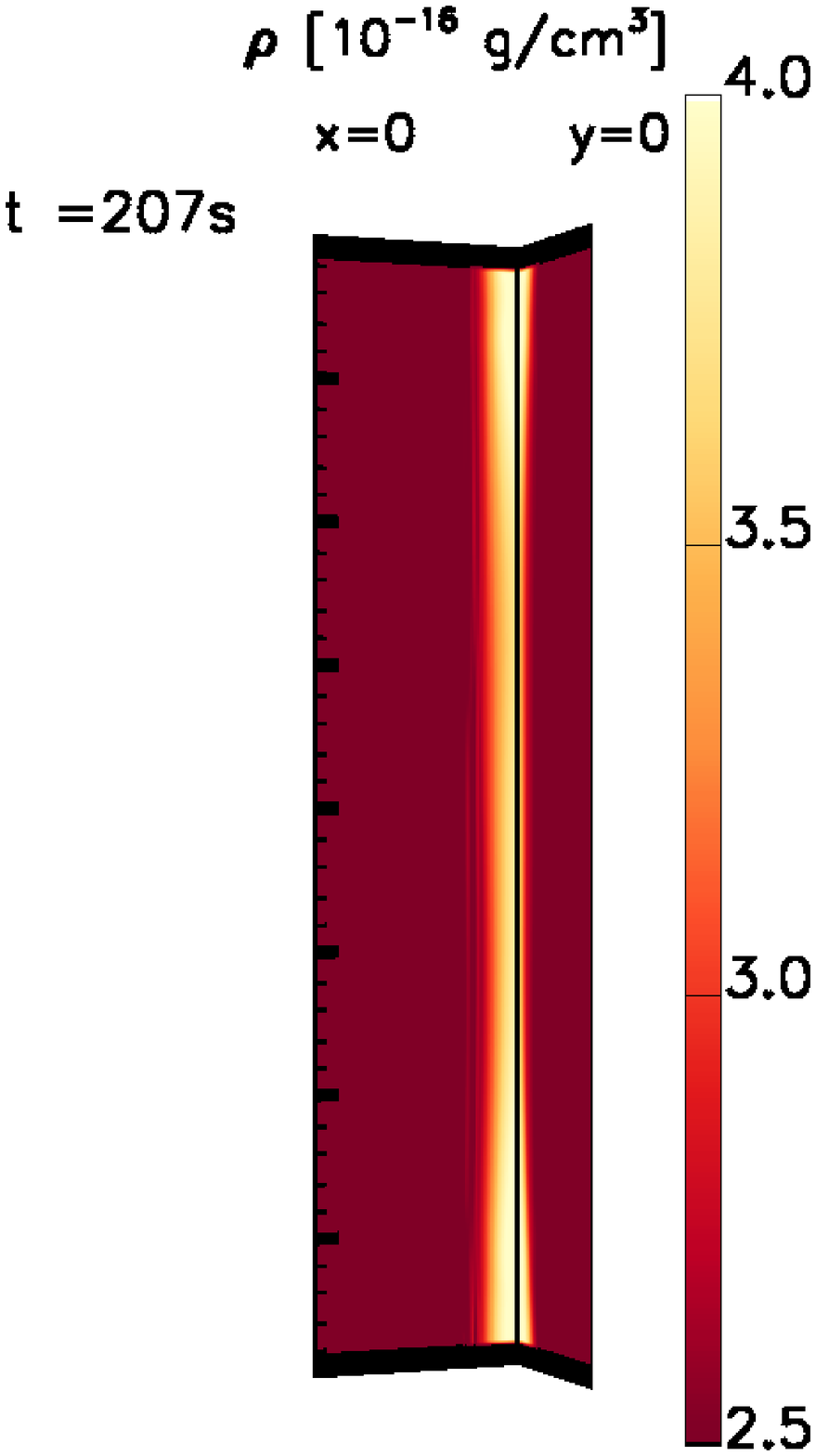} % printer
\includegraphics[scale=0.28,clip,viewport=110 50 380 650]{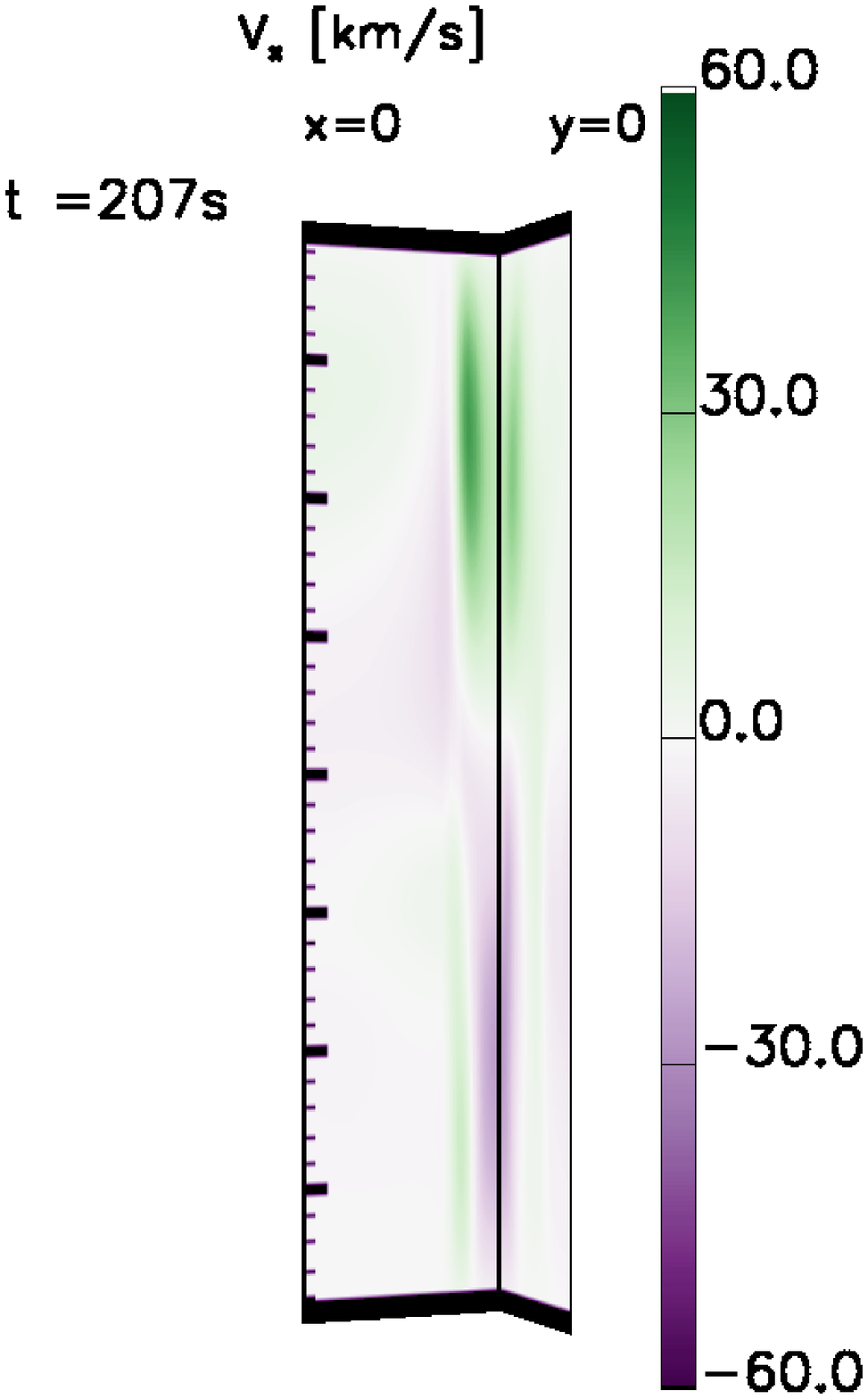} % printer
\includegraphics[scale=0.28,clip,viewport=110 50 290 650]{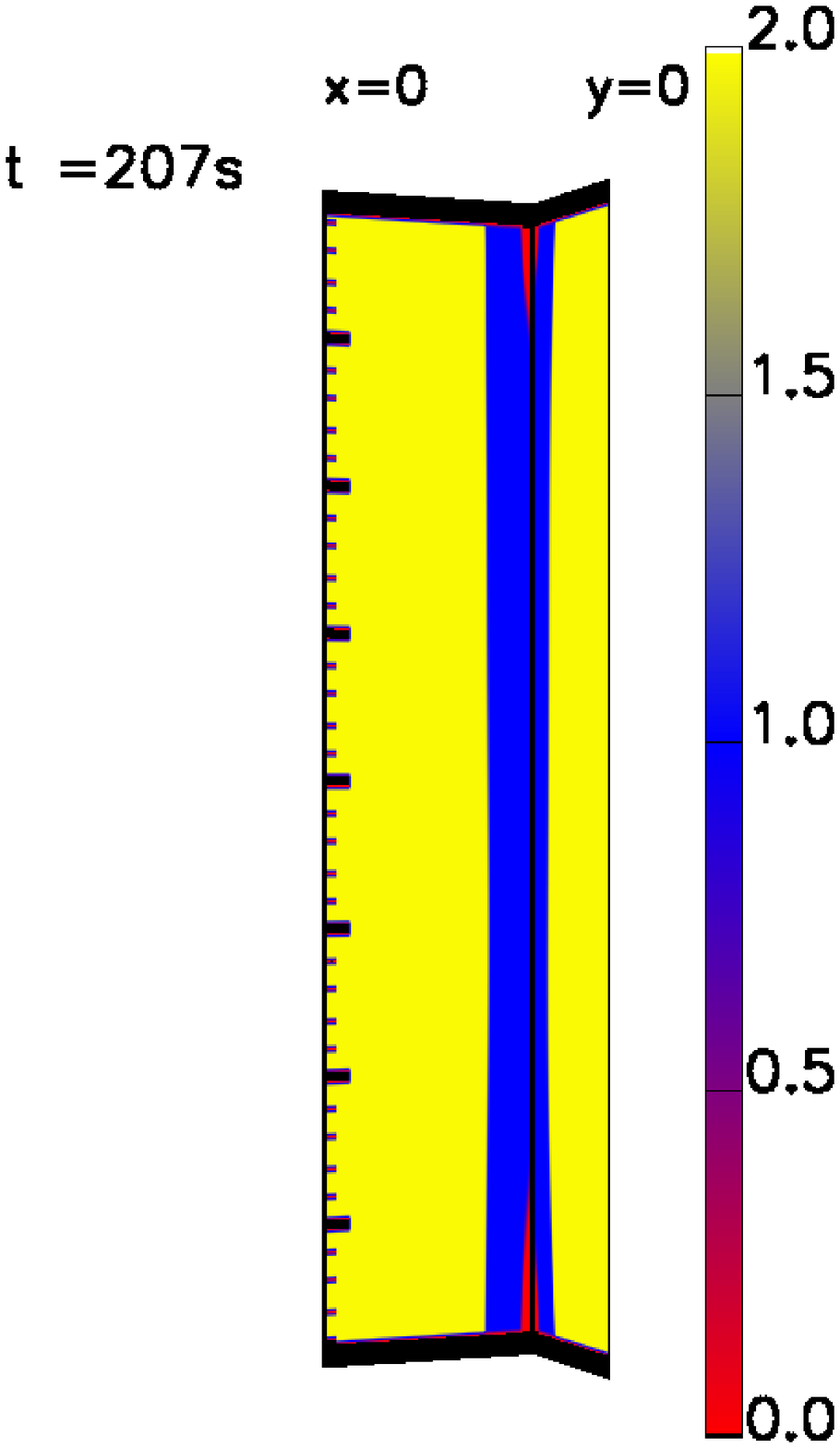} % printer
\caption{3D cuts of the MHD simulation domain showing maps of
density $\rho$ (left column),
$v_x$ (centre column),
and the tracers distribution (right column,
$tr_i$ is red, $tr_{bs}$ is blue, and $tr_e$ is yellow})
on the $x=0$ and $y=0$ plane
at $t=0$ (top row), $t=P/4$ (middle row), and $t=3P/4$ (bottom row).
\label{mhdevol}
\end{figure}

Figure~\ref{mhdevol} shows maps of density $\rho$, the
$x$-component of the velocity, and the tracers distribution 
at $t=0$, $t=P/4$, and $t=3P/4$
in a projection of the 3D simulation box where
we cut two vertical planes at $x=0$ and $y=0$
to show the initial oscillation of the flux tube.

\begin{figure}
\centering
\includegraphics[scale=0.32]{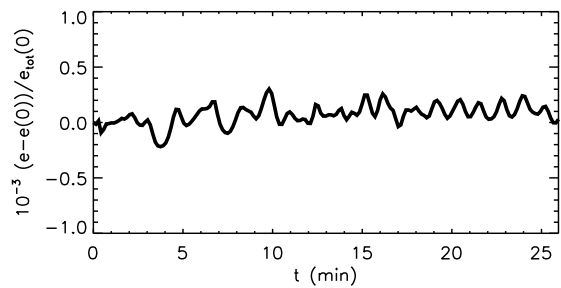}
%\caption{Energy difference normalised by the initial total energy as a function of time.}
\caption{Variation of total energy with time (normalised by its initial value).}
\label{totalenergy}
\end{figure}

We follow the evolution of the system over $5.5$ cycles of the fundamental mode, corresponding to approximately $26$ minutes.
The oscillation of the flux tube is damped and energy is converted into thermal energy.
We verify that energy is conserved within the simulation domain and find that total energy is conserved within $10^{-3}$ times its initial value (see Fig.~\ref{totalenergy}).

\subsection{Evolution of harmonics}
\label{dampmodes}

%We follow the standing oscillation of the flux tube for $5.5$ periods.
The damping of the kink oscillations occurs at different times for different modes due to the frequency-dependence of mode coupling \citep[independently of the damping regime, e.g. the Gaussian or exponential regimes explained in][]{2015A&A...578A..99P}.
To verify this, we use a minimization technique to find the best set of parameters to describe the oscillation, both as displacement and as a velocity perturbation.

First, we define the displacement of the flux tube
as a function of $z$ and $t$ using the centre of mass of the tracer of the interior region $tr_i$:
\begin{equation}
s_x(z,t)=\frac{\int tr_i(x,y,z,t) x dy}{\int tr_i(x,y,z,t) dy}.
\end{equation}
At any time $t$ we numerically find the coefficients $w_{s1}(t)$, $w_{s2}(t)$, and $w_{s3}(t)$
that minimise the following function with an accuracy
substantially better than the numerical resolution:
\begin{equation}
\label{chisequation}
\chi_s(t)=\int_{z_{min}}^{z_{max}} \left(s_x(z,t)-\left(w_{s1} f_{1}(z)+w_{s2} f_{2}(z)+w_{s1} f_{3}(z)\right) \right)^2 dz.
\end{equation}
In other terms we find the best fit for the function $s_x (z,t)$ in terms of the three parallel harmonics we initially excited.
The same method can be applied to the $x$-component of the velocity perturbation $u_x$ defined as:
\begin{equation}
u_x(z,t)=\frac{\int tr_i(x,y,z,t) v_x dy}{\int tr_i(x,y,z,t) dy}.
\end{equation}
To find $w_{u1}(t)$, $w_{u2}(t)$, and $w_{u3}(t)$ we minimise the function $\chi_u(t)$ with an accuracy
again substantially better than the numerical accuracy.

\begin{figure}
\centering
\includegraphics[scale=0.35]{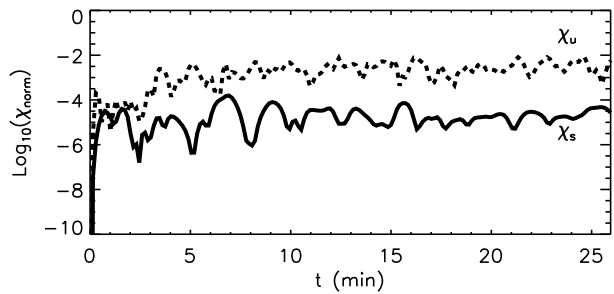}
\caption{Log of $\chi_s(t)$ (Eq.\ref{chisequation}) and $\chi_u(t)$ as a function of time
divided by the maximum value of the integral of function $s_x(z,t)^2$ and $u_x(z,t)^2$ along the $z$-direction, respectively.}
\label{chifigure}
\end{figure}

Figure~\ref{chifigure} shows the evolution of the function
$\chi_s(t)$ and $\chi_u(t)$ normalized by the maximum value of the integral of the functions $s_x(z,t)^2$ and $u_x(z,t)^2$ along the $z$-direction, respectively.

In Fig.~\ref{chifigure} the values of $\chi_s(t)$ and $\chi_u(t)$ remain about three orders of magnitude
smaller than the maximum of the functions $s_x(z,t)^2$ and $u_x(z,t)^2$.
This means that the error of representing the oscillation in terms of the first three harmonics is less than $1\%$.
This error is nearly absent at $t=0$ when the initial analytic profile can be reconstructed, but it then rises and remains constant.
Also, $\chi_u(t)$ remains about an order of magnitude larger than $\chi_s(t)$, as the velocity profile is more complex than the displacement.

\begin{figure}
\centering
\includegraphics[scale=0.28]{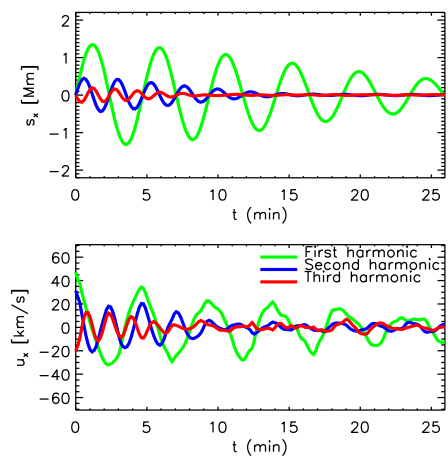}
\caption{Coefficients for the first three harmonics to fit
the displacement of the loop ($w_{s1}(t)$, $w_{s2}(t)$, and $w_{s3}(t)$ - upper panel)
and the velocity of the loop ($w_{u1}(t)$, $w_{u2}(t)$, and $w_{u3}(t)$ - lower panel) as a function of time.}
\label{wswv}
\end{figure}

Figure~\ref{wswv} shows the evolution of the coefficients 
$w_{s1}(t)$, $w_{s2}(t)$, and $w_{s3}(t)$ (upper panel) and 
$w_{u1}(t)$, $w_{u2}(t)$, and $w_{u3}(t)$ (lower panel).
Analysing the oscillation either from the point of view of the displacement or velocity perturbation, it is evident that the three parallel harmonics are damped in time and this occurs at different rates.
The third harmonic (red line) is the the most strongly damped, followed by the second harmonic (blue line), and finally the first harmonic (green line), which remains measurable and is not completely damped by the end of the simulation.
The second and third harmonics take between $4$ and $5$ cycles to be damped.
The oscillations in $u_x$ are significantly more noisy
than those in $s_x$.

The different damping times of the three harmonics is explained by their different periods. %and the frequency-dependence of mode coupling.
%In fact, in our configuration t
The kink oscillation harmonics are damped as they transfer energy to Alfv\'en modes in the cylinder boundary and the efficiency of this mode coupling depends linearly on the period of the kink mode \citep[e.g. thin tube approximation by][]{RudermanRoberts2002}.
If we define the damping time ($\tau_n$) of the $n$th harmonic as the time when its maximum oscillation displacement is reduced to about a third of the initial one, we find that
$\tau_1=1480$~s, $\tau_2=600$~s, and $\tau_3=355$~s.
These values slightly depart from a linear dependence,
however we note that i) our flux tube departs from the thin flux tube approximation
and ii) we solve non ideal MHD equation and iii) and our initial kink oscillations is weakly non-linear.
Thus our numerical experiment does not entirely satisfy the approximations of analytical approaches such as \citet{RudermanRoberts2002}.
% the best fit is tau=tau_3*(period/period_3)^1.3

%\begin{figure}
%\centering
%\includegraphics[scale=0.28]{figure/energies_3phpi_tr_x8_y10_nzb_fll_.ps}
%\caption{}
%\label{wswv}
%\end{figure}
%questo potrebbe mostrare che l'energia termica smette di salire quando l'energia cinetica si e' dissipata
%anche se non si vede che l'una sale mentre l'altra scende
%perche' la trasformazione e' mediata dal campo magnetico che non 
%ha un'evoluzione molto comprensibile qua.
%quando inizia a salire e' per il riscaldamento al peak della velocita'
%(it should be checked if in the simulation with only one harmonic, it starts later)
%%print,total((wu1+wu2-wu3)^2)=4.6278375e+15
%%print,total((wu1)^2)=2.8424461e+15
%%maybe I should run the simulation with more energy in the w1 mode

%- to commment on the offset between the centre and the boundary shell
%- to derive KHI conditions

\subsection{Heating}
\label{heating}

%[maybe we find that the region involved in heating does not correspond to the tracers]

\begin{figure}
\centering
\includegraphics[scale=0.35]{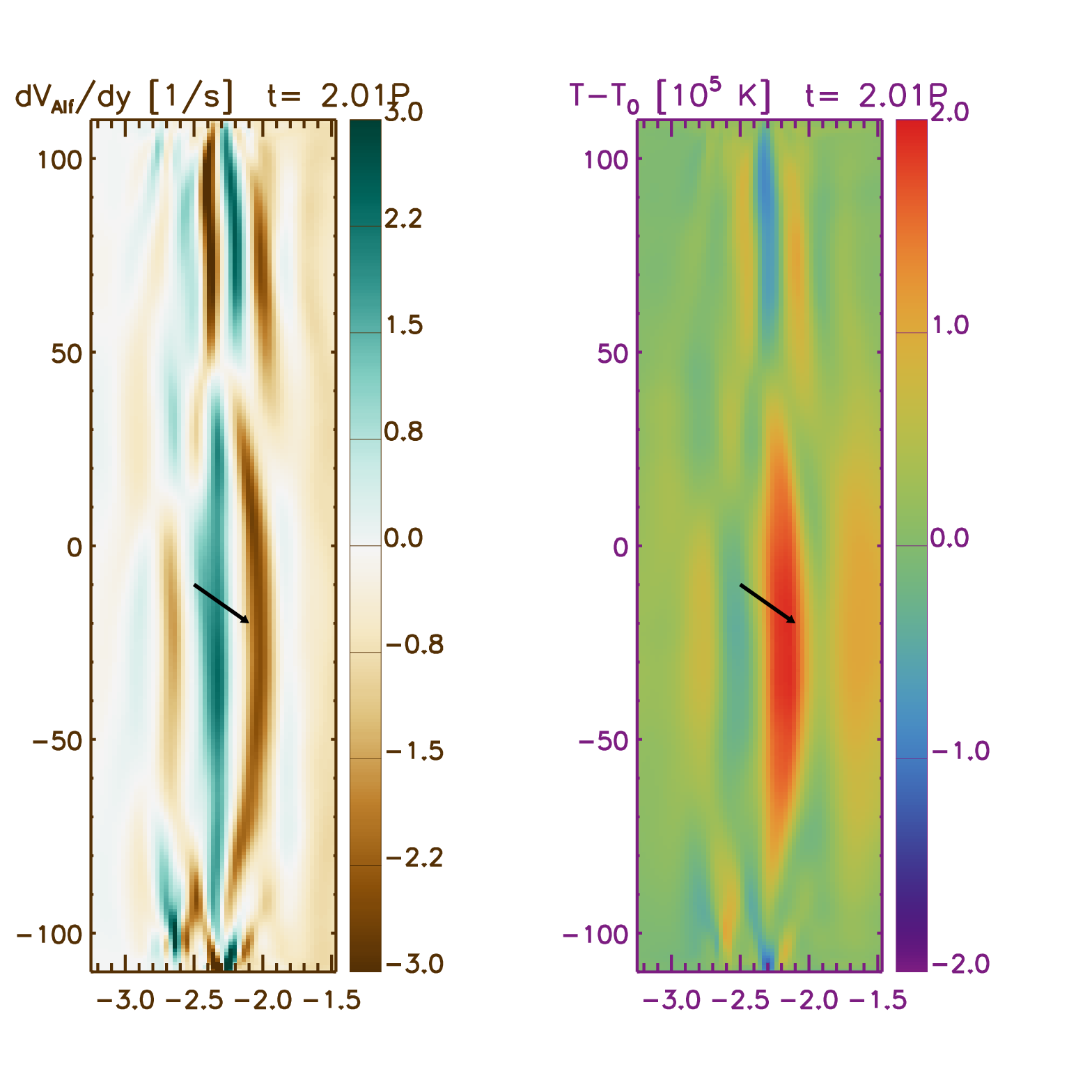}
\includegraphics[scale=0.35]{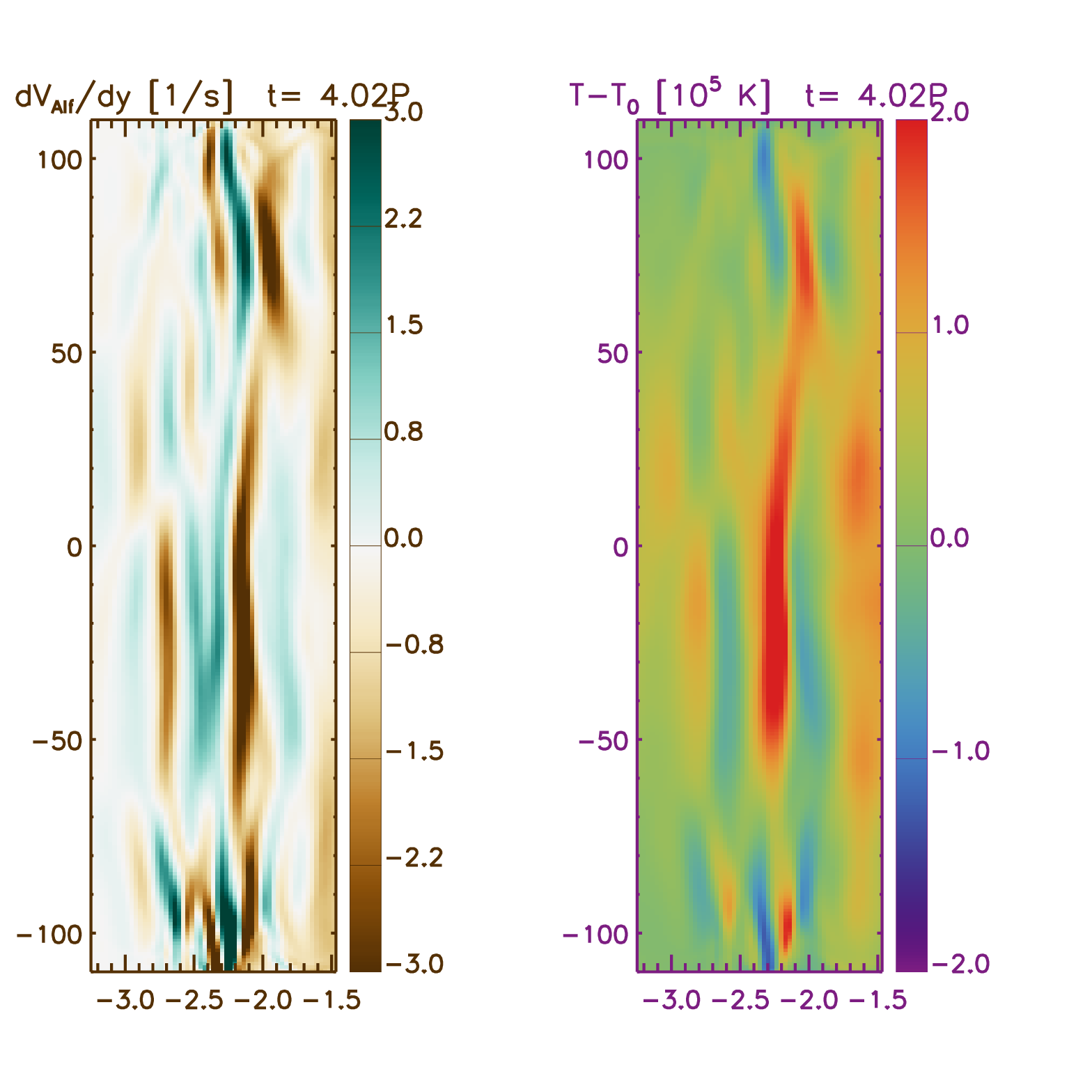}
\caption{Maps of Alfv\'en speed gradients (left column) and temperature increase (right column)
at $t=2P$ (top panels) and $t=4P$ (bottom panels).
The arrows in the top panels show where the phase-mixing initially develops and leads to a temperature increase.}
\label{dalfhaet}
\end{figure}

Our model naturally leads to the phase-mixing of Alfv\'en waves generated by mode coupling from the
initial kink oscillation of the flux tube.
Therefore, the temperature mostly increases 
where the Alfv\'en speed has steepest gradients,
as larger inhomogeneities favour the phase-mixing rate.
In Fig.~\ref{dalfhaet} we show simultaneously the gradient of Alfv\'en speed and the temperature variation from the initial condition in a portion of the $x=0$ plane at $t=2P$ and $t=4P$.

We find that at any time there is an evident correlation between the location of steep Alfv\'en speed gradients and the temperature increase, as the patterns are very similar in the two maps.
Additionally, we also find that both Alfv\'en speed gradient and temperature variation patterns lack symmetry about the $z=0$ plane because of the initial asymmetric kink oscillation.
The phase-mixing quickly develops (top panels) near $z=-20$~Mm where
the initial peak of the kink velocity is located,
as the Alfv\'en speed gradient is large (either negative or positive) and the temperature increases significantly.
At later stage (bottom panels) the heating is more pronounced near the centre of the flux tube after the second and third harmonics have been dissipated.

Therefore, we find that the plasma heating connected to the phase-mixing of Alfv\'en waves primarily occurs where the oscillations are stronger and as the Alfv\'en speed varies only across the boundary shell, we can focus on this specific tracer to investigate this plasma heating process.
This approach means that we focus on the heating due to the actual change of temperature of the plasma and our analysis is not
affected by apparent heating such as plasma mixing
(due to KHI as in Fig.\ref{horcuts}).

\begin{figure}
\centering
\includegraphics[scale=0.35,clip,viewport=60 0 655 330]{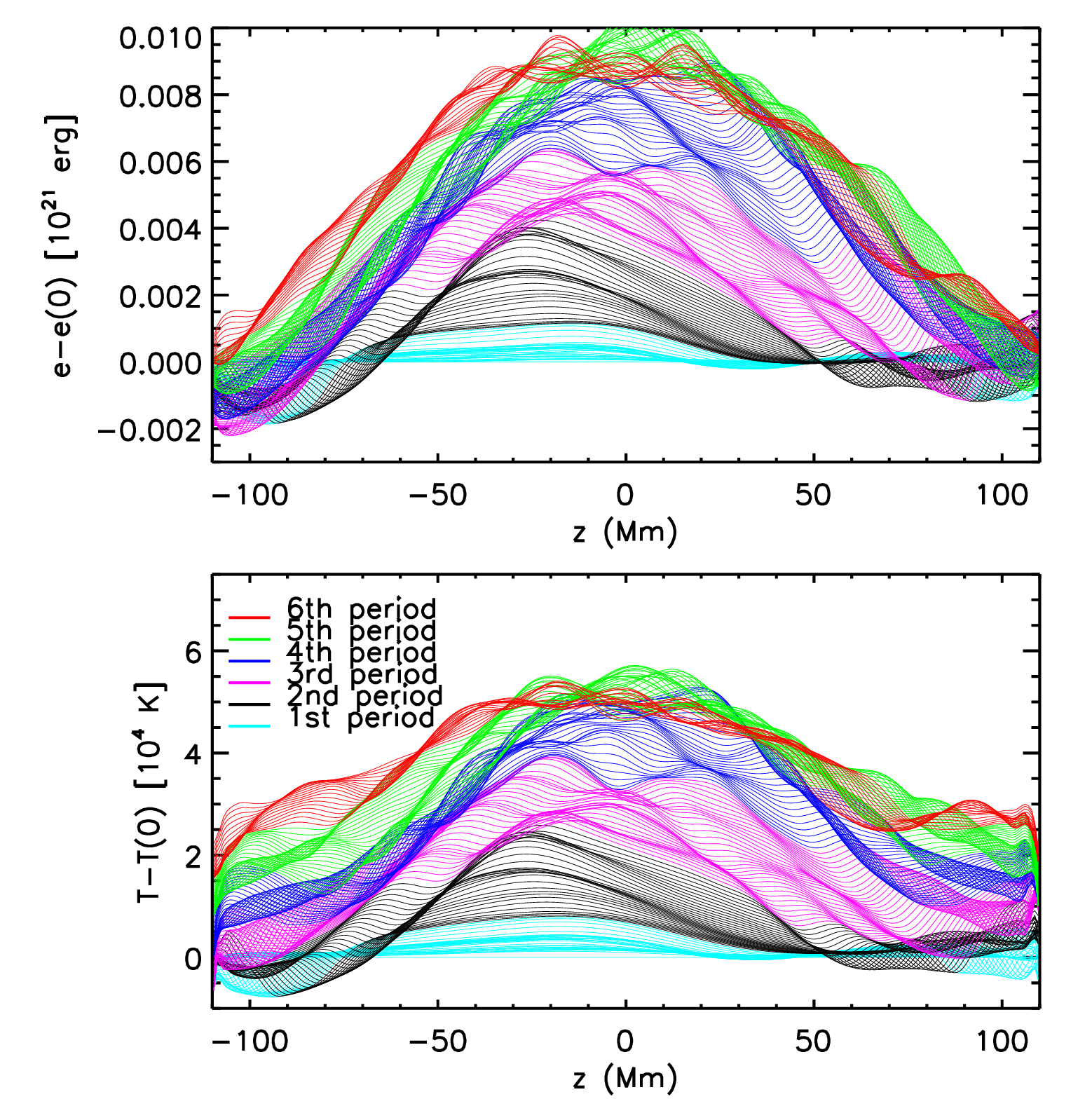}
\caption{Average temperature increase for the boundary shell tracer $tr_{bs}$ as function of $z$ at different times.
Different colours denote different oscillation cycles.}
\label{heatingz}
\end{figure}

In Fig.~\ref{heatingz} we plot the average temperature increase for the boundary shell tracer $tr_{bs}$ as a function of the coordinate $z$ at various times,
where we use different colours for subsequent oscillation cycles.

The temperature starts to increase during the first few cycles (light blue and black lines)
near $z=-20$~Mm, where the initial velocity peak is located.
However, in the subsequent cycles we find a higher temperature increase near $z=0$, as the oscillatory motion is dominated by the first harmonic at this stage.
Finally, for the last couple of cycles (green and red lines) we see that the temperature increase is saturated near $z=0$ as the lines overlap.
At the same time the distribution of temperature increase becomes less peaked and more distributed around the centre.
The timescale of this change in the temperature distribution is comparable with the timescale of thermal conduction to thermalise the plasma over approximately $25$~Mm.
However, we conclude that the presence of additional harmonics influences the location of the heating as the interference between different harmonics and the different damping times lead to a non-stationary velocity distribution along the flux tube as the velocity peak drifts in time.

\begin{figure}
\centering
\includegraphics[scale=0.35,clip,viewport=60 0 655 330]{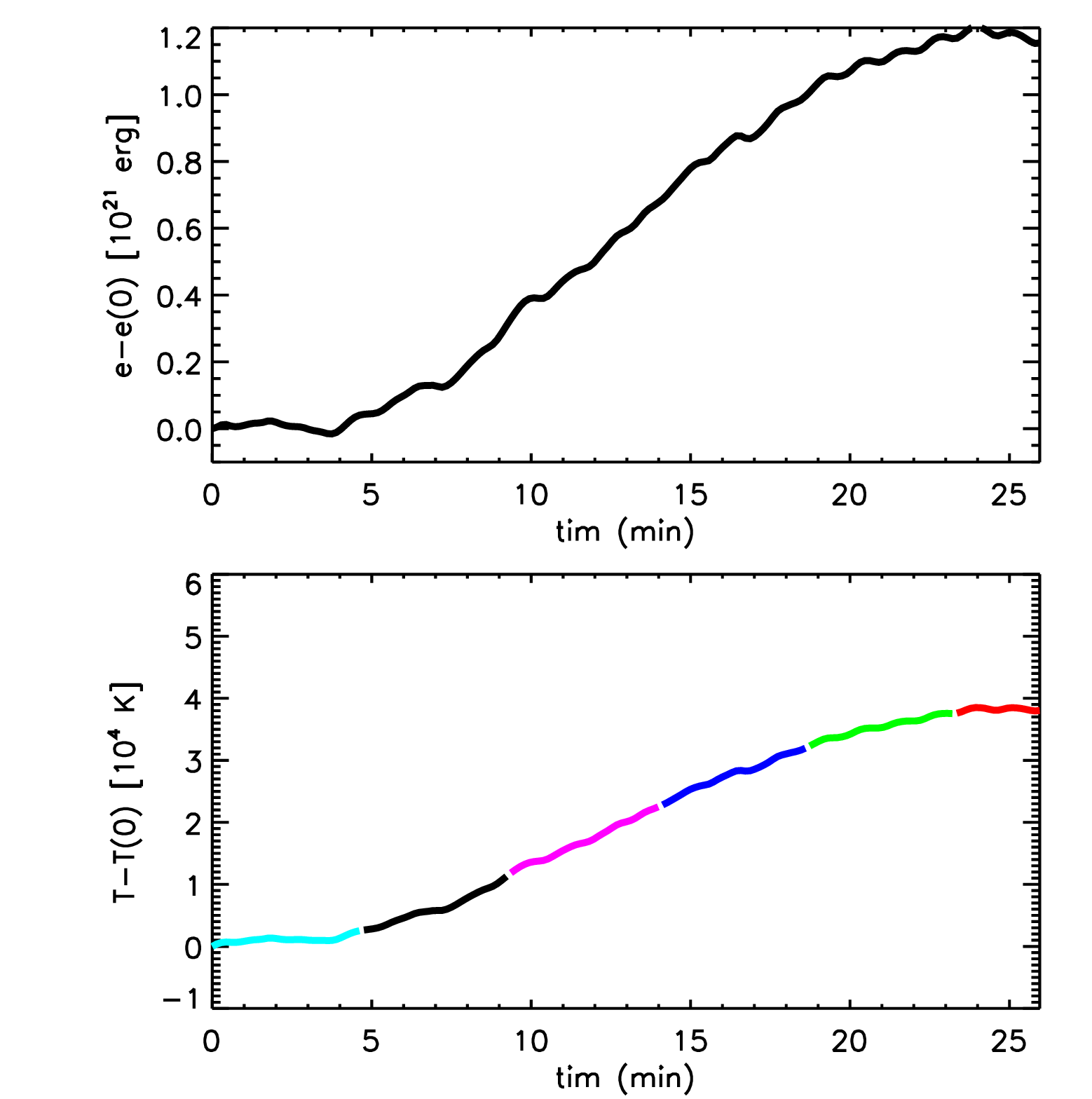}
\caption{Average temperature increase of the boundary shell tracer as a function of time.
The colour table represents different oscillation cycles as in Fig.~\ref{heatingz}.}
\label{heatingt}
\end{figure}

A similar behaviour is found on a global scale, i.e. averaging the plasma temperature variation for the boundary shell tracer across the whole domain,
as shown in Fig.~\ref{heatingt}.
We see that the temperature remains unaltered for the first $4$ minutes (corresponding to the first cycle) and then it starts to increase roughly linearly in time.
After $20$ minutes (approximately $5P$), the profile starts to saturate and the temperature increase plateaus at about $40000$~K.
%From an energy point of view,
This temperature increase corresponds to a very modest
deposition of thermal energy in the plasma.
A simple estimation of radiative losses \citep{Rosner1978} over the same time span shows that the energy radiated by the plasma would be about three orders of magnitude larger
than the energy deposited by heating.

\begin{figure}
\centering
\includegraphics[scale=0.28,clip,viewport=440 10 1320 690]{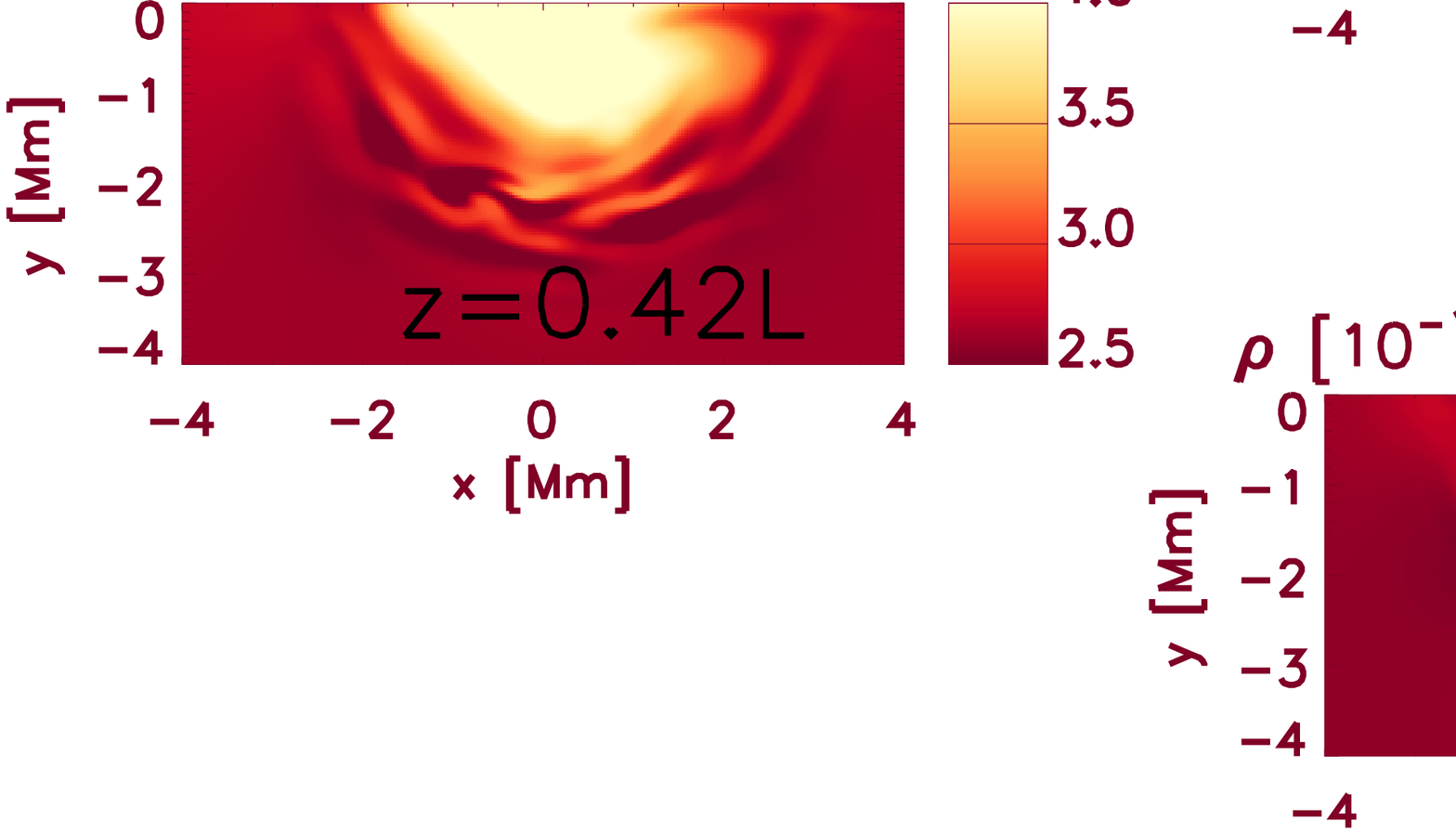}
\includegraphics[scale=0.28,clip,viewport=440 10 1320 690]{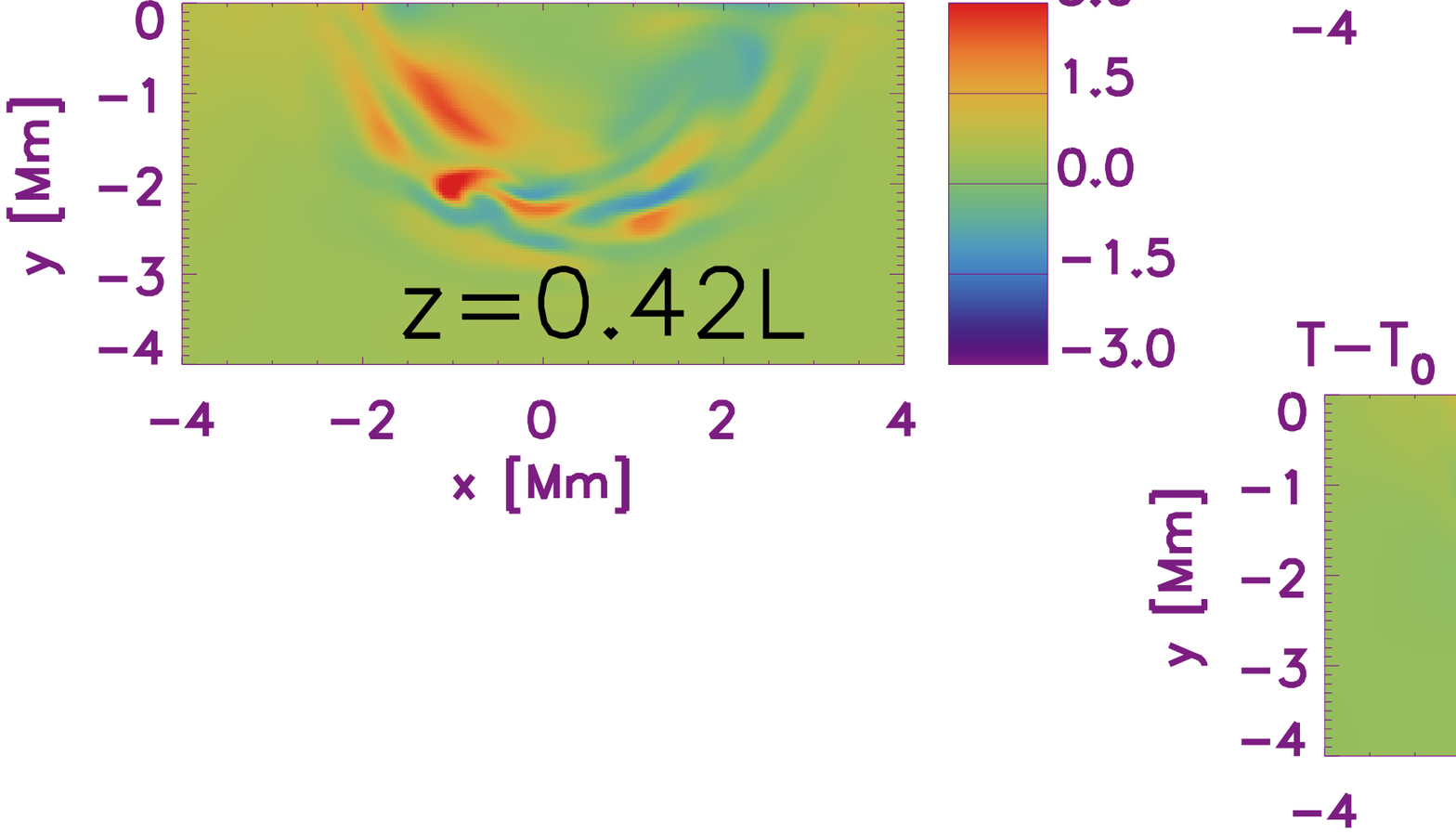}
\caption{Cross sections of density (upper panels)
and temperature variation (lower panels)
at $t=5P$ for different locations.
First column: anti-nodes of the second harmonic.
Second column: anti-nodes of the third (and first) harmonic.}
\label{horcuts}
\end{figure}

At the same time, the structure of the flux tube is modified by the development of smaller scale deformations on the boundary layer due to the Kelvin-Helmholtz instabilities (KHI) generated by the azimuthal motions \citep[e.g][]{Terradas2008,Antolin2015,Howson2017}.
In Fig.~\ref{horcuts} we show the development of these structures at $t=5P$ on some characteristic cross sections, namely the anti-nodes of the second harmonic (first column) and the anti-nodes of the third harmonic (second column).
As expected, the KHI structures lead to a partial 
fragmentation of the flux tube structure and some mixing of plasma.
We also find that the KHI structures do not develop symmetrically about the apex ($z=0$) owing to the asymmetric oscillation.
As expected, they are more developed on the half of the flux tube where the initial plasma velocity is higher.
By comparing the patterns in the cross sections of density and temperature we find that the regions of increased temperature match with those where the KHI are also more developed, and that the oscillatory motion displaces high temperature regions out of the phase-mixing plane ($x=0$).

%  if you plot w3(boundary shell trace)-w3(loop tracer) vs time it behaves like a beat where
%  cos(om3*t)+cos(om2*t)=cos((om3+om2)t/2)cos((om3-om2)t/2)
%  with om2 and om3 that are the frequencies corresponding to the periods of second and third harmonics
% this is probably not so interesting

\subsection{Fundamental mode simulation}

In order to further address how additional parallel harmonics influence the flux tube oscillation dynamics and the following heating we perform a simulation in which only the fundamental mode of the kink oscillation is excited.
In this simulation we trigger the initial oscillation with the same kinetic energy that
we have initially input in the simulation described in Sect.~\ref{referencesim}, 
i.e. $6.7\times10^3$ $ergs$.
In this way we aim to determine how the heating deposition is affected by the different harmonics while keeping the initial energy budget unchanged.

\begin{figure}
\centering
\includegraphics[scale=0.28]{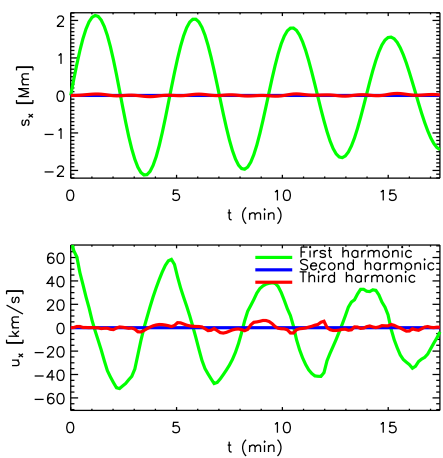}
\caption{Coefficients for the first three harmonics to fit
the displacement of the loop and the velocity of the loop as a function of time for the simulation with only the fundamental mode excited by the initial perturbation}.
\label{wswv1st}
\end{figure}

Figure~\ref{wswv1st} summarises this simulation, showing the time dependent evolution of the fundamental mode
(as in Fig.~\ref{wswv}).
We find that the oscillation undergoes more modest damping, and that the second and third harmonics are not excited (as expected).
In this simulation, the damping of the oscillation is slower
than in our reference simulation because the fundamental mode has slower damping times than the higher harmonics whose energy is more quickly transferred to Aflv\'en modes.

\begin{figure}
\centering
\includegraphics[scale=0.35,clip,viewport=60 0 655 330]{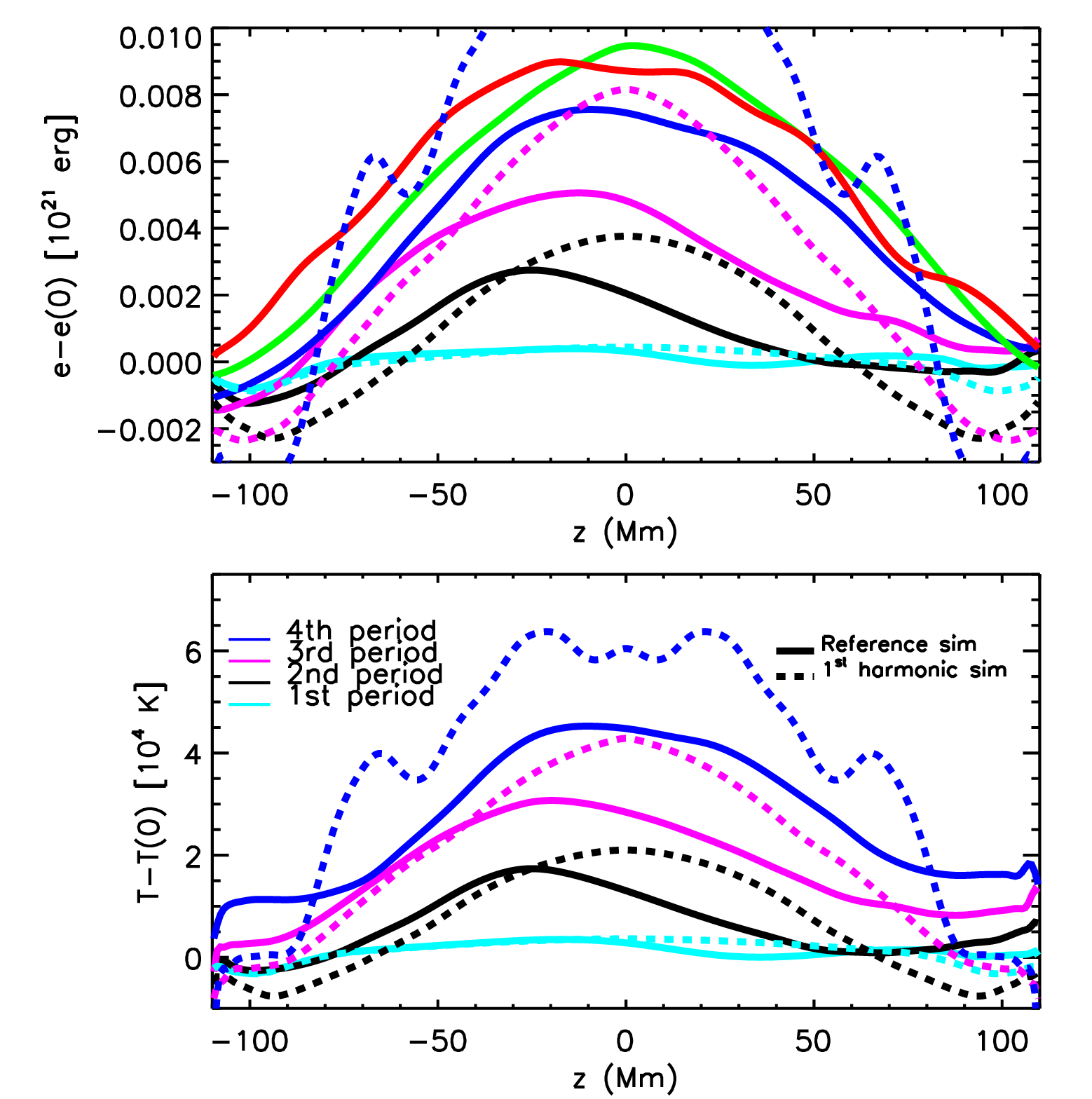}
\caption{Temperature difference as a function of $z$ averaged over one period.
Continuous and dashed lines correspond to simulations with three harmonics or the fundamental mode only, respectively.}
\label{heatingzcomparison_1st}
\end{figure}

Moreover, the distribution of the temperature increase is affected by the different velocity profile.
Figure~\ref{heatingzcomparison_1st} shows the temperature increase (for the boundary shell tracer, thus addressing only actual plasma
temperature changes) averaged over one period time span for subsequent cycles and compared for the two simulations.
We find that the temperature increase peaks at $z=0$
in the simulation with only the fundamental mode and it remains symmetric about $z=0$.
For the first couple of cycles, there seems to be no difference in the total temperature increase,
although the heating occurs at different locations.
However, from the third cycle the temperature increases 
more when we excite only the fundamental mode.

\begin{figure}
\centering
\includegraphics[scale=0.35,clip,viewport=60 0 655 330]{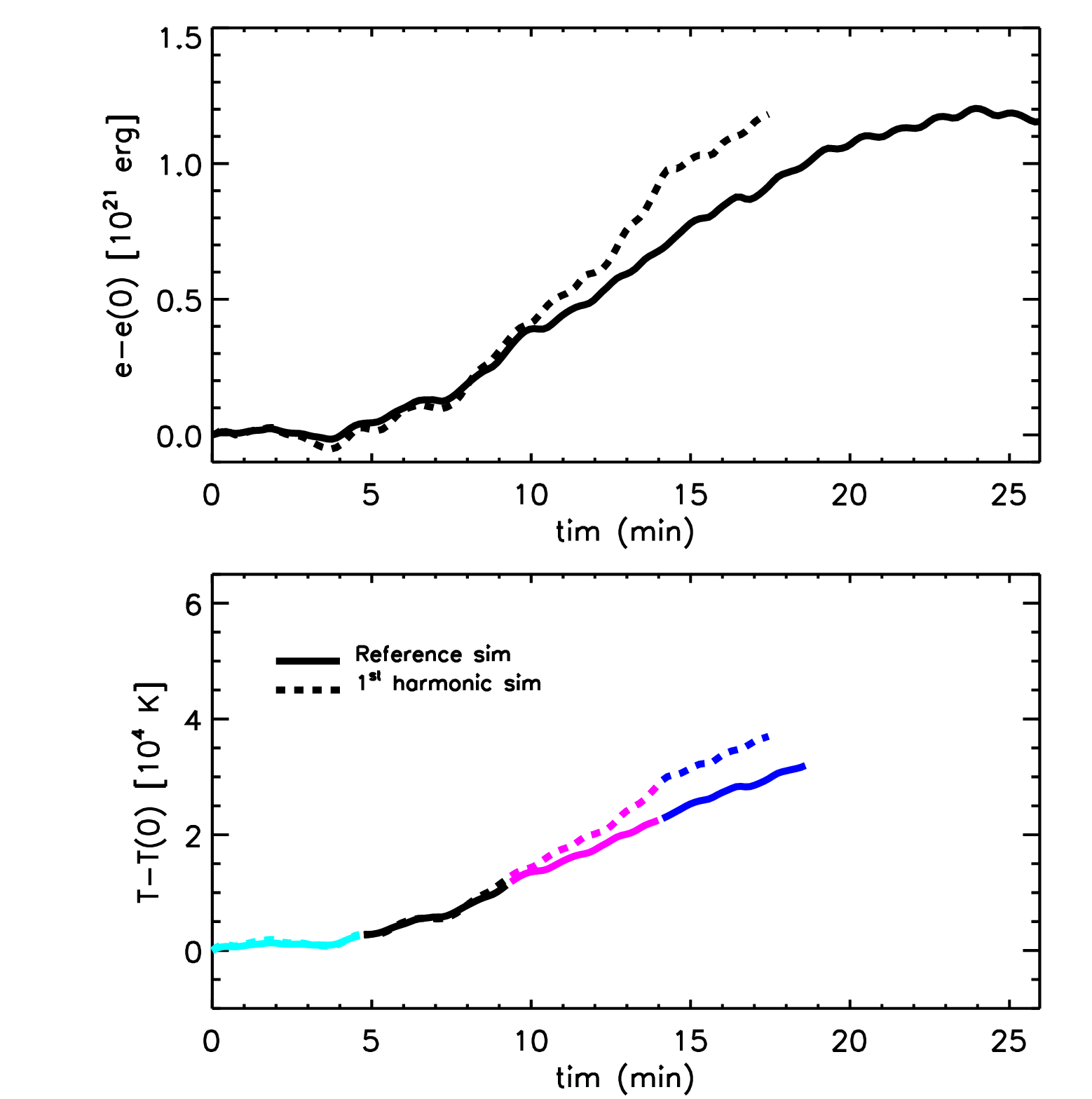}
\caption{Average temperature increase of the boundary shell tracer as a function of time.
The colour table represents different oscillation cycles as in Fig.~\ref{heatingz}.
The continuous line represents the simulation with additional harmonics, while the dashed line represents the simulation with the fundamental mode only.}
\label{heatingtcomparison_1st}
\end{figure}

This is more evident if we plot the averaged temperature increase over the entire numerical domain as function of time (Fig.~\ref{heatingtcomparison_1st}).
We find that the temperature increases in a very similar way for the first couple of periods and then the 
simulation where only the fundamental harmonic is excited shows a temperature increase about $20\%$ larger than the reference simulation.

\begin{figure}
\centering
\includegraphics[scale=0.55]{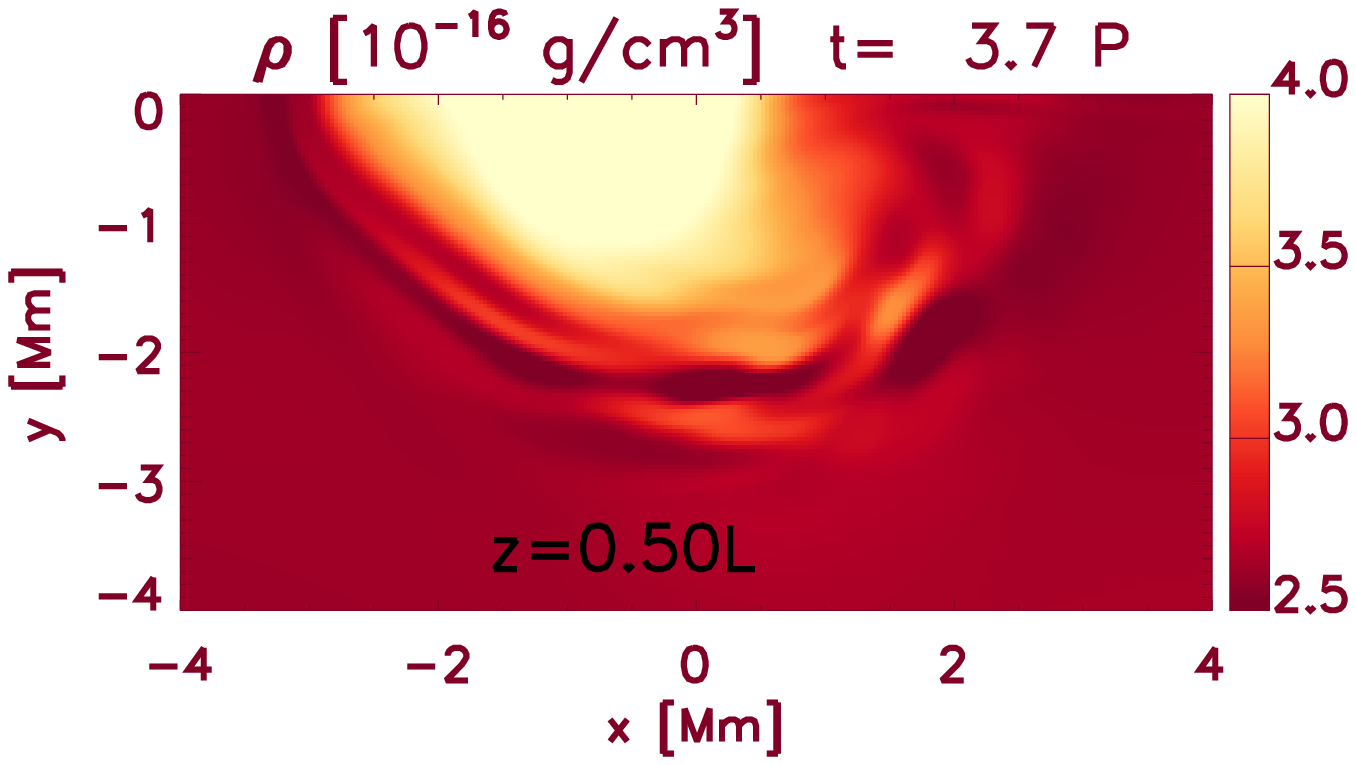}
\includegraphics[scale=0.55]{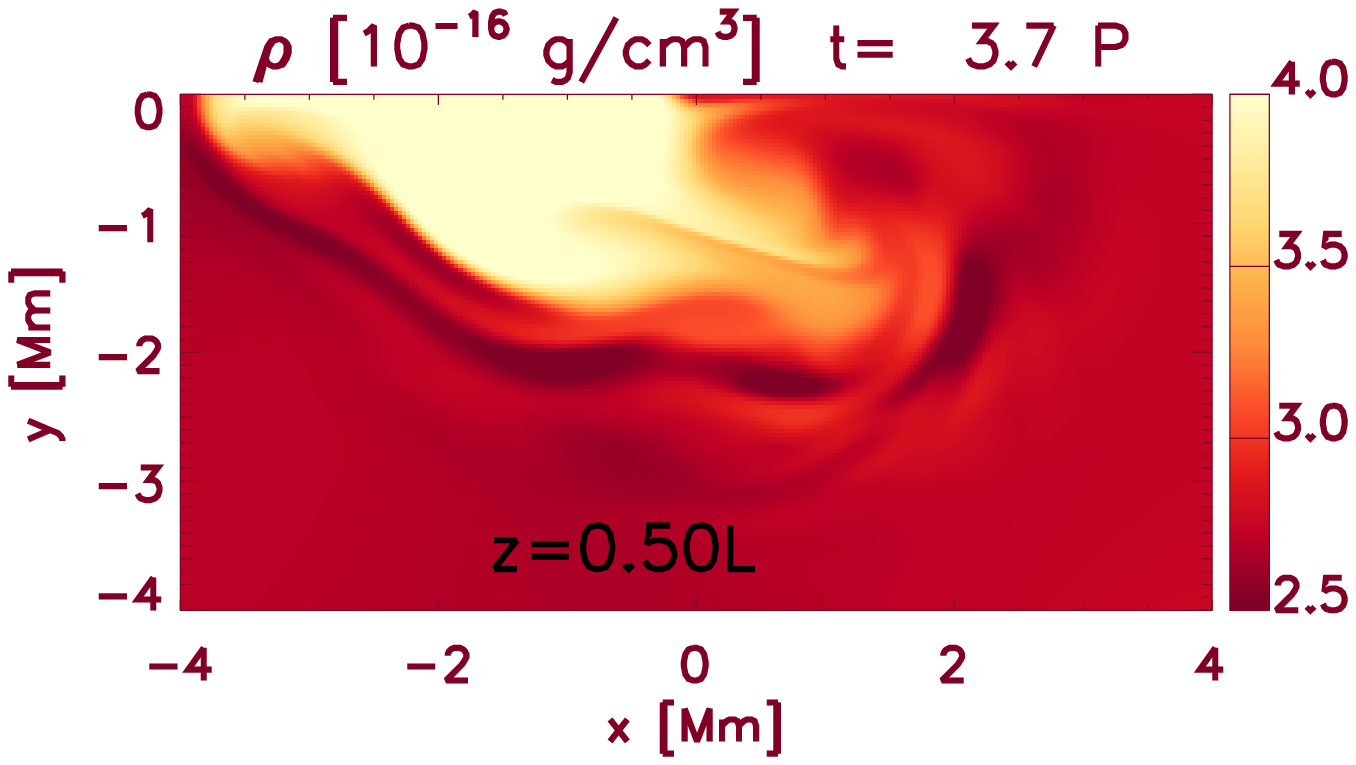}
\caption{Comparison of the cross sections of the density for the simulations with 3 harmonics (left panel) and with 1 harmonic (right panel) at the centre of the flux tube $z=0$ at $t=3.7$ $P$ that shows the different development of the KHI in the two simulations.}
\label{horcuts1stmode}
\end{figure}

This can be explained by the different initial velocity profiles and how they affect the development of KHI.
Figure~\ref{horcuts1stmode} shows the cross section of density at $z=0$ and $t=3.7P$ for the two simulations.
The simulation with only one harmonic shows more developed KHI than the one with three harmonics.
This also implies that when only the fundamental harmonic is excited, smaller scales develop faster and thus the heating developing from KHI triggers earlier.
This can account for the different temperature increase evolution.
When only the fundamental harmonic is excited, KHI are generated more efficiently because the damping of this mode by mode coupling is slower 
than for higher order harmonics and thus higher velocity shears can persist longer.
Additionally, the interference between the three harmonics is not always constructive and at times this can reduce the shear velocity between the oscillating flux tube and its surroundings. A coherent fundamental mode oscillation is not affected by that.

\subsection{Boundary shell}

We also consider the role of the width of the boundary shell in the phase-mixing heating process.
We perform an additional simulation with a boundary layer half the size of that %$\epsilon=0.575$,
used in the simulation presented in Sect.~\ref{referencesim}.

As discussed in \citet{PaganoDeMoortel2017},
the width of the boundary shell plays two competing 
roles in this setup.
On one hand, the efficiency of the mode coupling from kink oscillations to Alfv\'en waves is proportional to the width of the boundary shell.
On the other hand the efficiency of the phase-mixing mechanism is inversely proportional to the width of the boundary shell.
Thus the energy deposition is not significantly affected by the width of the boundary shell,
while the temperature increase is slightly affected,
as roughly the same amount of energy is deposited in a smaller volume.

\begin{figure}
\centering
\includegraphics[scale=0.35,clip,viewport=60 0 655 330]{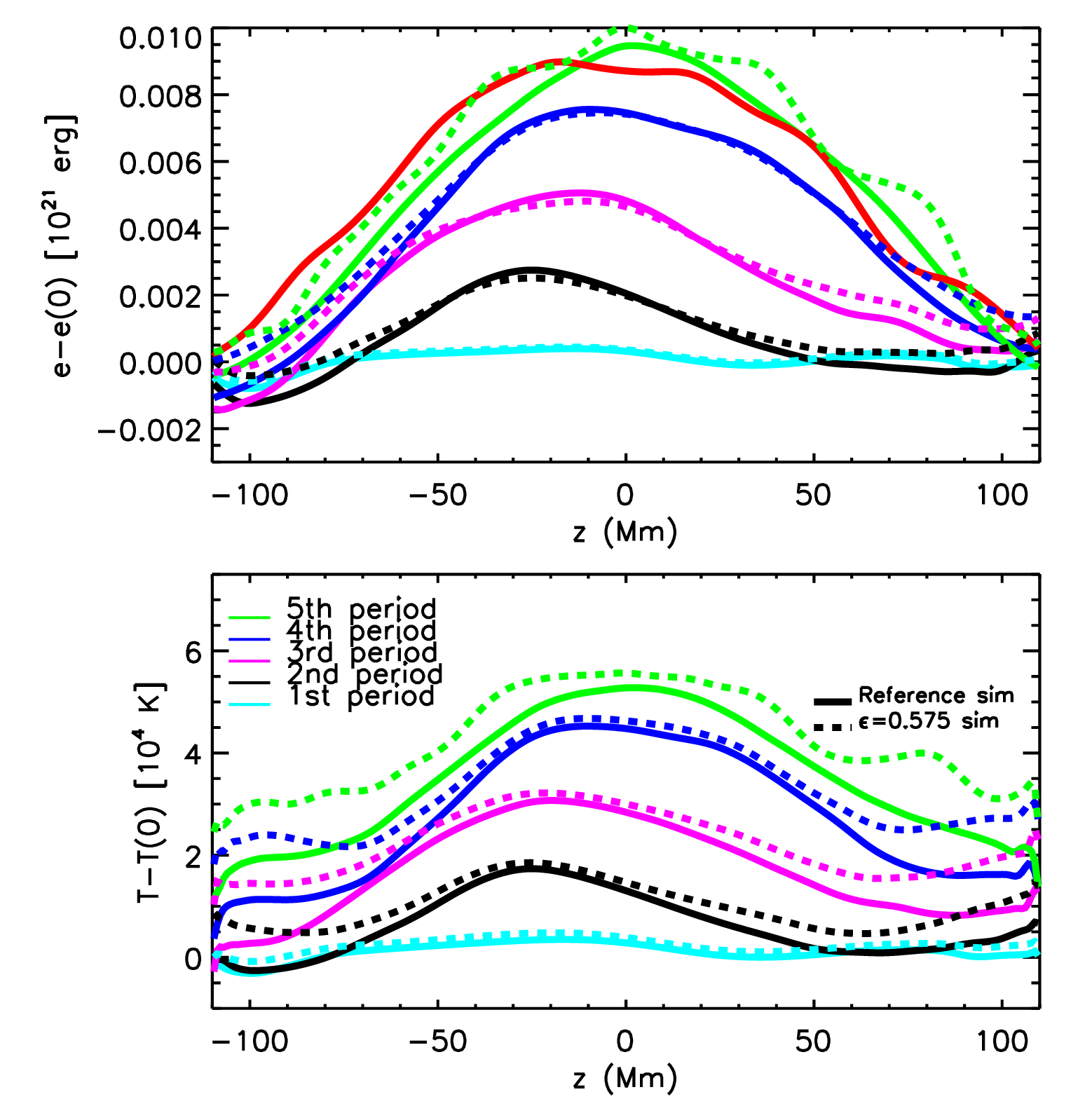}
\caption{Temperature change with respect to the initial value
as a function of $z$ averaged over one period oscillation for subsequent 
oscillations identified with different colors. 
Continuous line for simulation with $\epsilon=1.15$ 
and dashed line for simulation with $\epsilon=0.575$}
\label{heatingzcomparison}
\end{figure}

This behaviour is also confirmed in this study. 
In Fig.~\ref{heatingzcomparison}, we plot the
temperature variation (for the boundary shell tracer, thus addressing only actual plasma
temperature changes) along the flux tube averaged over each cycle.
We find that the two simulations behave very similarly,
and only develop a significant difference after $5$ cycles,
when the small scales generated by KHI are more diffused.

\begin{figure}
\centering
\includegraphics[scale=0.35,clip,viewport=60 0 655 330]{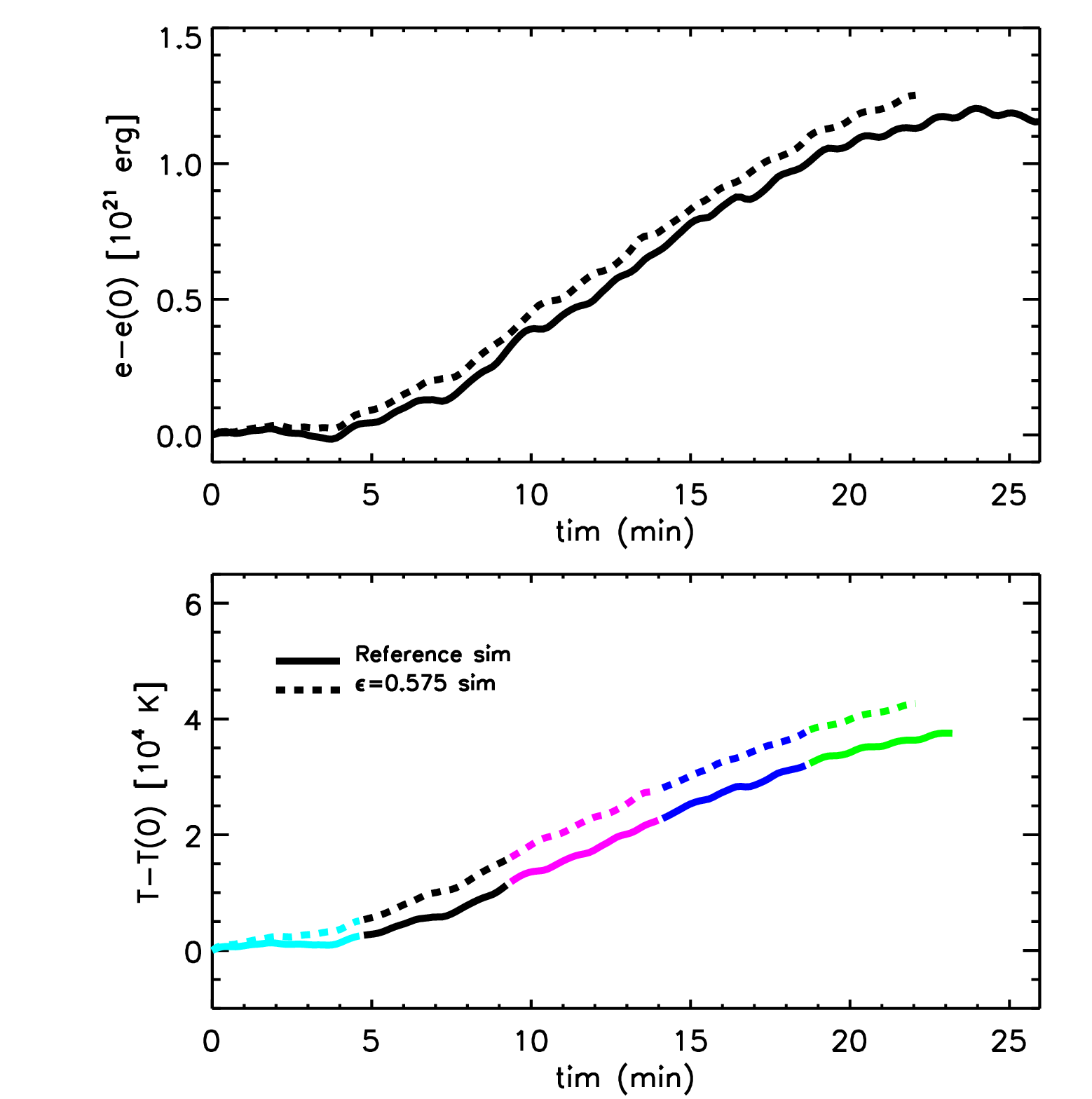}
\caption{Average temperature  change with respect to the initial value of the boundary shell tracer plasma as a function of time.
The colour table represents different oscillation cycles as in Fig.~\ref{heatingz}.
The continuous and dashed lines represent simulations with thicker and thinner boundary shells, respectively.}
\label{heatingtcomparison}
\end{figure}

By following the temperature increase averaged over the whole domain (Fig.\ref{heatingtcomparison})
we find that the two simulations show a very similar temperature increase, with the thinner boundary simulation showing only a $2\%$ larger temperature increase.
%A thinner boundary shell leads to higher temperature because, while the energy deposited by the mode-coupling and then phase-mixing is not dependent on the width of the boundary shell, the temperature does slightly, as the same thermal energy is deposited in a smaller region.
The energy deposited by the mode coupling and then phase-mixing is not dependent on the width of the boundary shell. However, a thinner boundary shell leads to higher temperatures since the same thermal energy is deposited in a smaller region.

\section{Comparison with observations}
\label{compobs}
 
In this section, we analyse our simulation data using the 
seismological techniques from \citet{Pascoe2017}.
These techniques are based on the analysis of EUV imaging data. We may use our simulation data to generate the forward-modelled EUV intensity for a particular instrument and bandpass such as SDO/AIA $171${\AA} used in the observational analysis.
However, for the purpose of this study we use a simpler approximation of calculating the intensity as $\rho^2 T$ integrated along the line of sight ($y$ coordinate).
We create a time-distance (TD) map for $z=-22$~Mm, which corresponds to the position $0.4L$ and is comparable to the position of the observational slit used for Loop \#1 in \citet{Pascoe2016b}.
This allows us to generate a TD map which may be used in the same observational analysis routines used for observational data (but preserving the greater spatial resolution provided by our numerical data).

\begin{figure}
\resizebox{\hsize}{!}{\includegraphics{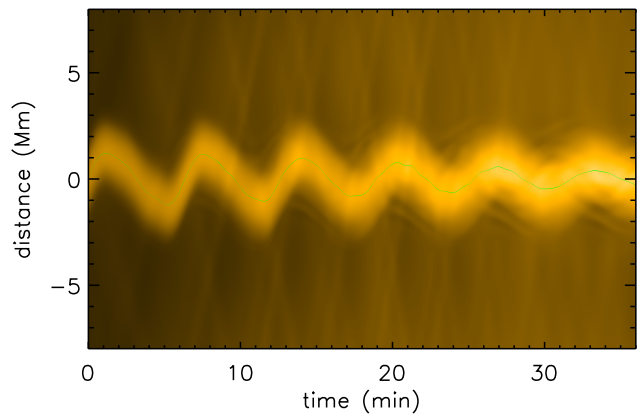}}
\caption{Time-distance map of the quantity $\rho^2 T$ integrated along the LOS at $z=0.4L$.}
\label{tdmap}
\end{figure}

\begin{figure*}
\centering
\includegraphics[width=8.5cm]{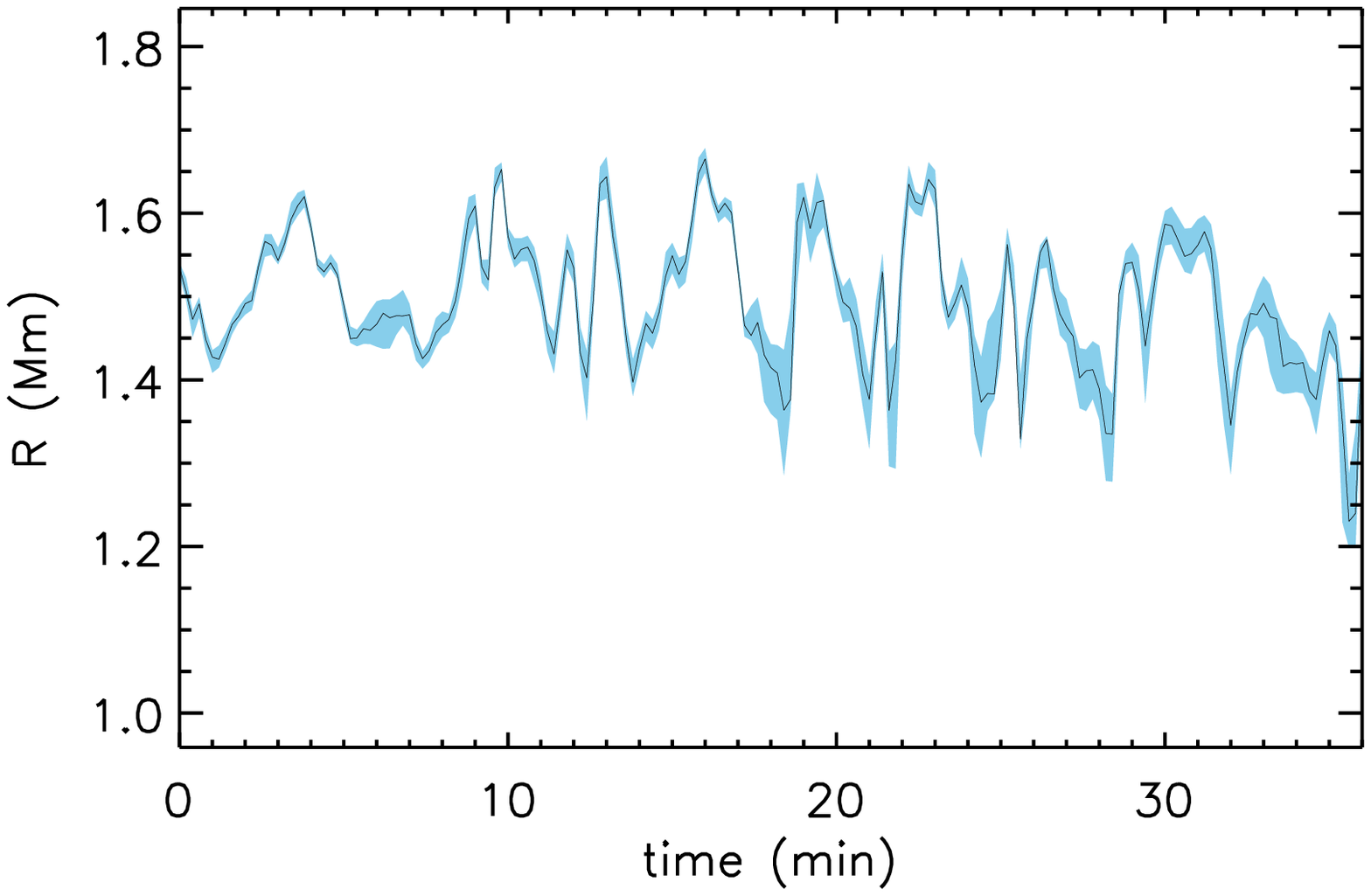}
\includegraphics[width=8.5cm]{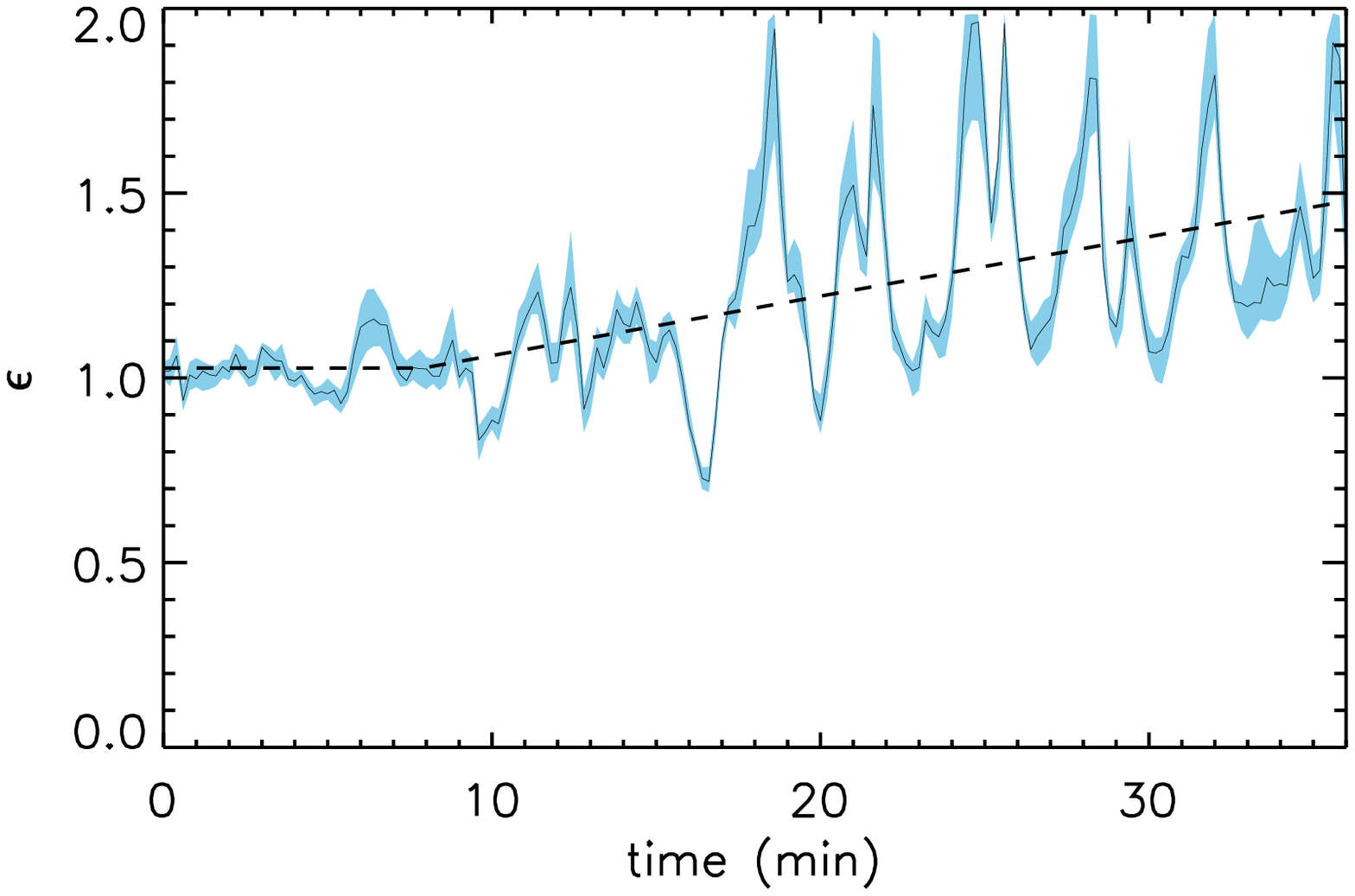}
\includegraphics[width=8.5cm]{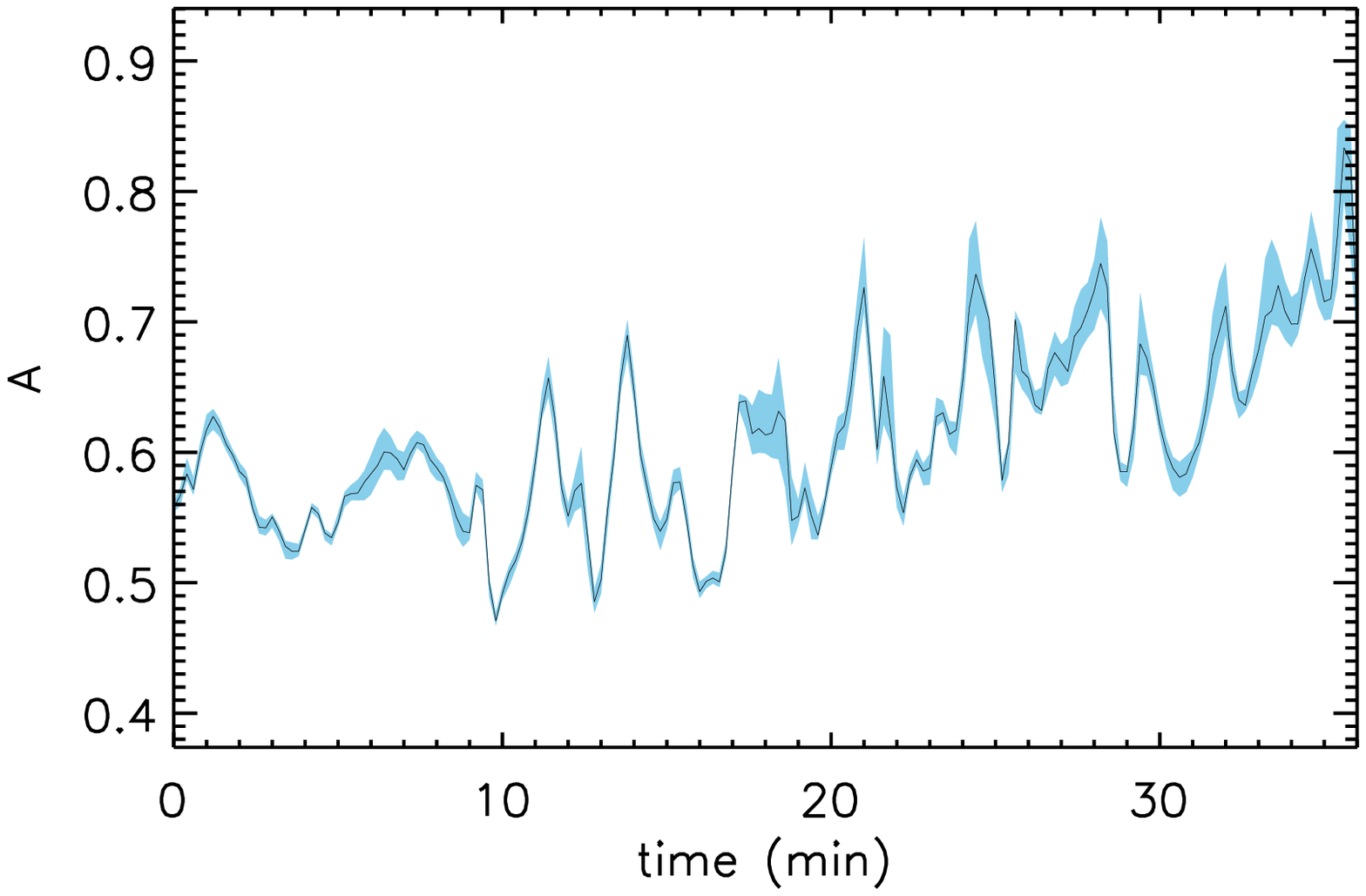}
\includegraphics[width=8.5cm]{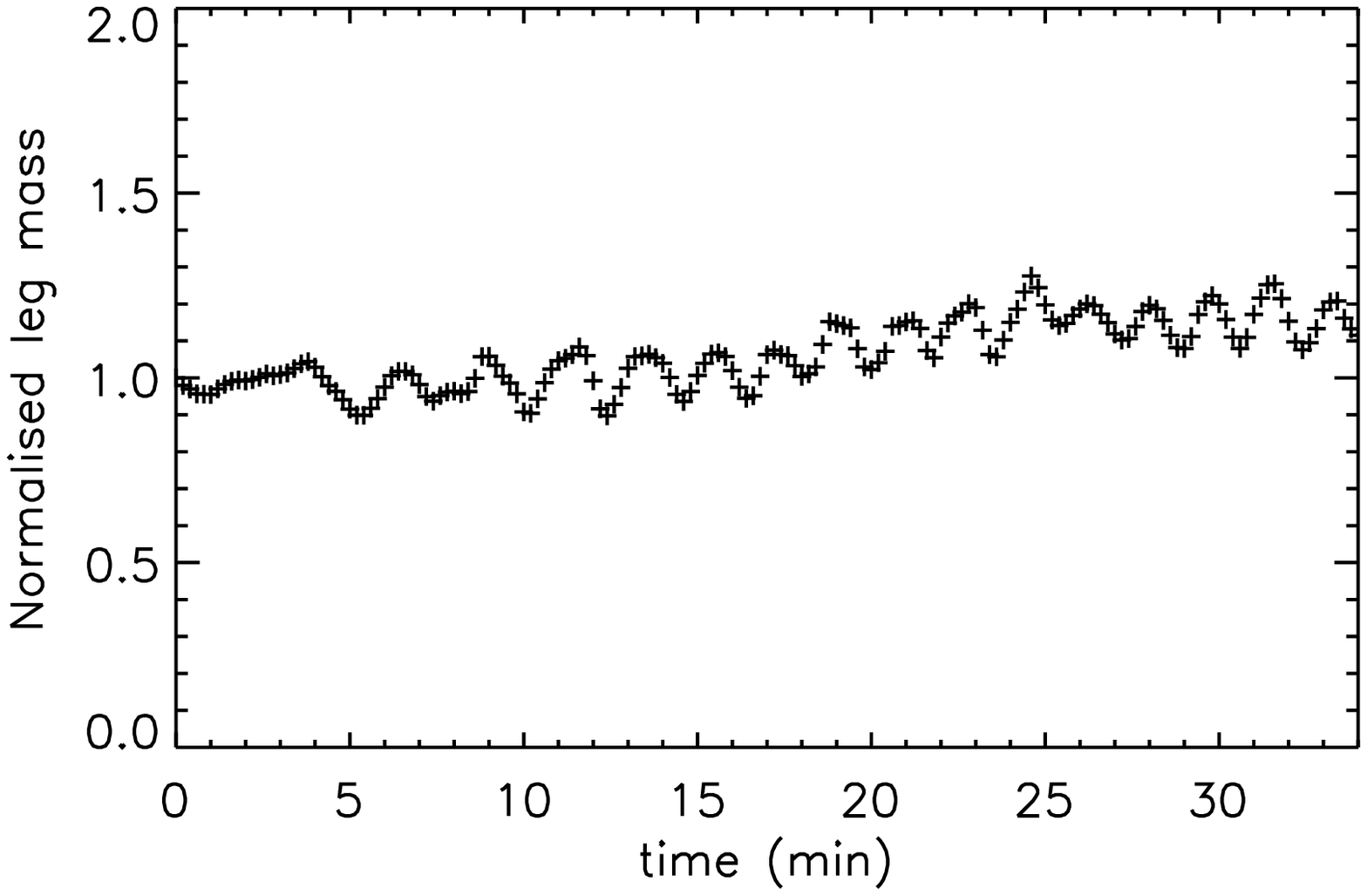}
\caption{Time-dependence of the transverse density profile parameters $R$ (top left), $\epsilon$ (top right), and $A$ (bottom left) estimated by forward modelling of the transverse intensity profile.
The symbols indicate the maximum a posteriori probability values and the shaded regions correspond to the 95\% credible intervals.
The bottom right panel shows the variation in leg mass per unit length.}
\label{ft}
\end{figure*}

Figures~\ref{tdmap} and \ref{ft} show the results of estimating the transverse density profile by forward modelling of the EUV intensity profile \citep{2017A&A...600L...7P,2017A&A...605A..65G}
(we note that the effect of the point spread function is not considered).
Figure~\ref{tdmap} shows the TD map with the position of the loop shown by the green line (corresponding to the maximum a posteriori probability value returned by MCMC sampling).
Figure~\ref{ft} shows the variation of $R$, $\epsilon$, $A$, and the normalised leg mass (per unit length) during the oscillation.
We find a gradual increase in $\epsilon$ after about a period of oscillation.
These results are consistent with \citet{goddard_khi} who find the signatures of KHI on a forward modelled loop core to include a widening inhomogeneous layer and an approximately stationary radius. However, in contrast to that study, we find an increasing rather than decreasing intensity, associated with the loop heating.
Localised KHI structures are also responsible for the rapid changes in $\epsilon$ values (whereas our fit is based on a single loop without this fine structuring).
We also note that Fig.~\ref{tdmap} shows additional waves outside the loop corresponding to propagating fast waves. These can contribute to the variation of density profile parameters with a short periodicity determined by the small size of the numerical domain in the directions transverse to the loop axis.
Also \citet{Antolin2017} found that the onset of KHI leads to the apparent broadening of the boundary shell in forward-modelled observations.

There is also a gradual increase of approximately 10\% in the leg mass by the end of the simulation and a short period oscillation that begins at the same time as the KHI onset time. This is consistent with KHI generating fine structuring in the inhomogeneous layer which affects the estimate of the monolithic loop profile used by the forward modelling method (and based on the linear transition layer profile).

\begin{figure}
\centering
\resizebox{\hsize}{!}{\includegraphics{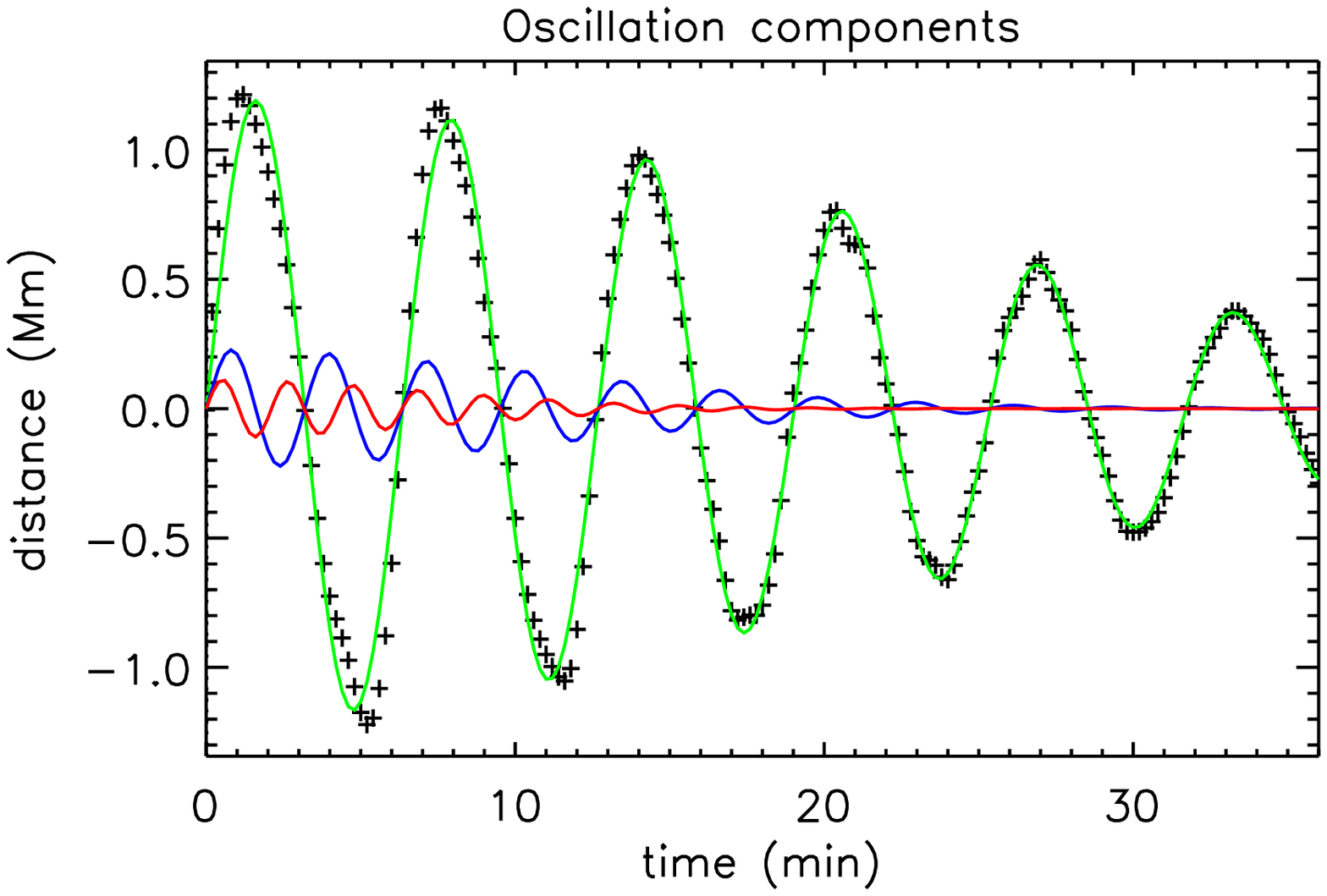}}
\resizebox{\hsize}{!}{\includegraphics{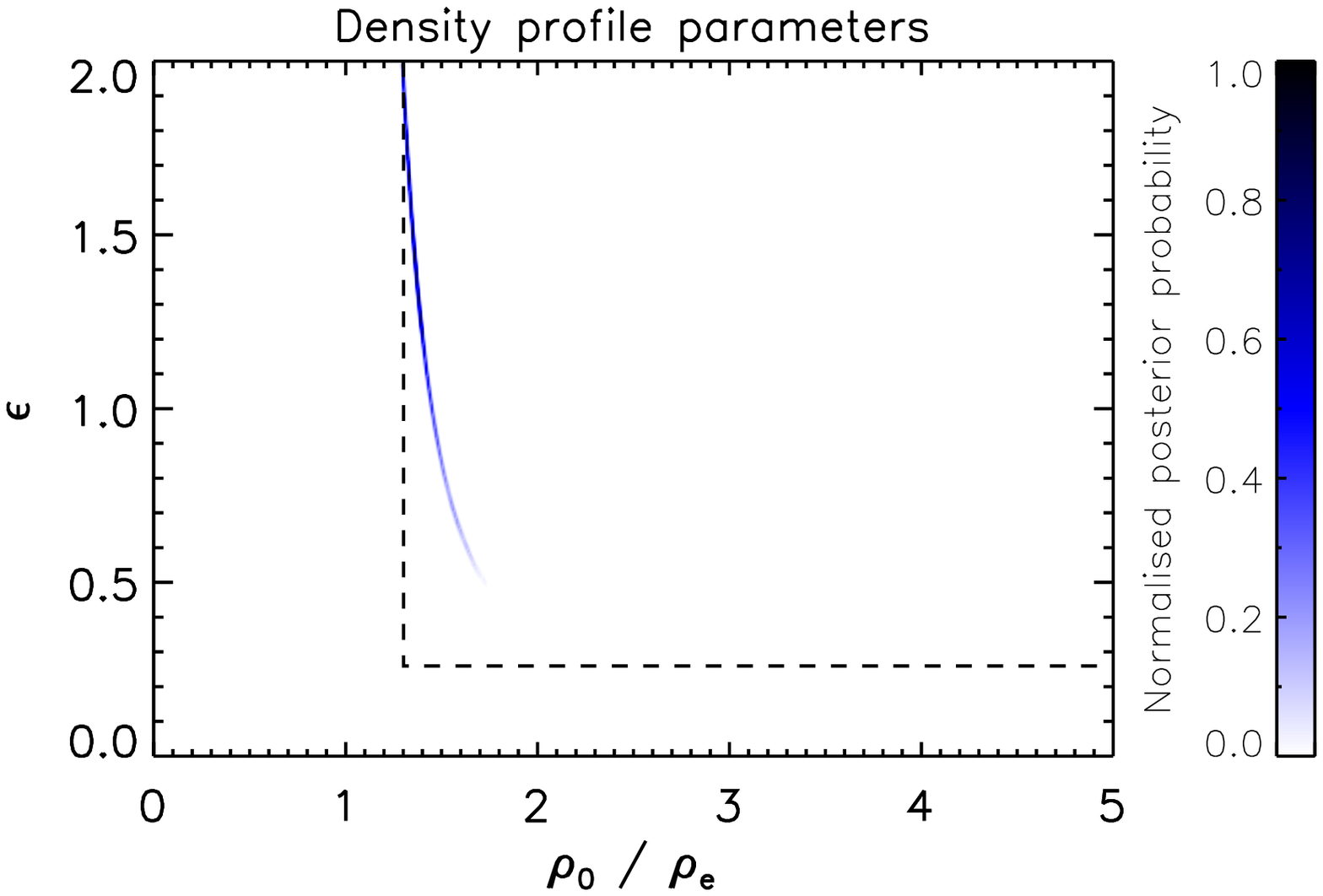}}
\caption{Results of the seismological analysis of the loop oscillation.
The top panel shows the detrended loop position (symbols) with the first (green), second (blue), and third (red) parallel harmonics.
The bottom panel shows the density profile parameters determined by the oscillation damping envelope.
The dashed lines correspond to estimates for the lower limits of the density contrast ratio and inhomogeneous layer width.}
\label{seismology}
\end{figure}

%From this analysis we derive the displacement of the flux tube as a function of time and we thus derive the intensity of the 1st, 2nd and 3rd harmonics.
%This leads to harmonics amplitude evolution and decay compatible with the one actually found in the simulation and described in Sect.~\ref{referencesim}.

Figure~\ref{seismology} shows the results of the seismological analysis of the kink oscillation.
These panels may be compared with Fig.~6 of \citet{2017A&A...600A..78P}, with which there is good agreement.
The dashed lines correspond to limiting approximations based on the Gaussian damping profile being dominant for low density contrasts and the exponential damping profile for high density contrasts \citep{Hood2013,Pascoe2013}.
As in the case of the observational data, the oscillation being well-described by a Gaussian damping profile implies a loop with a low density contrast ratio.
This provides an estimate of the upper limit of $\rho_\textrm{0}/\rho_\textrm{e} \lesssim 2$, in addition to the lower limit prescribed by the overall damping rate.
On the other hand, for the low density contrast regime the inhomogeneous layer width is weakly constrained by the seismological method to $\epsilon \gtrsim 0.5$ due to the asymptotic behaviour \citep[e.g.][]{2007A&A...463..333A}.
The estimated density contrasts are slightly lower
than the density contrast of $\rho_c=1.7$
used in the MHD initial condition.
However, this method is based on the thin boundary approximation, whereas our simulation is for $\epsilon \sim 1$.
The difference of approximately $20\%$ is consistent with the study of the effect of large inhomogeneous layers on the resonant absorption damping time (for the exponential damping regime) performed by \citet{2004ApJ...606.1223V}.

\section{Discussion and conclusions}
\label{conclusion}

In this work we have carried out MHD simulations to study
how the excitation of multiple parallel harmonics of standing kink oscillations in coronal loops
influences the heating associated with the phase-mixing of Alfv\'en waves.
The work is motivated by the observation of
multi-harmonics oscillations in \citet{2017A&A...600A..78P}.
Our modelling focuses on a magnetised flux tube anchored at its foot points, where an initial velocity field is set to trigger 
the oscillation of the first three kink oscillation harmonics.
The properties of the flux tube and the initial velocity distribution
are based on a particular kink oscillation observed in \citet{2017A&A...600A..78P}, namely a loop with a low density contrast ratio, a wide inhomogenous layer, and observed to support three standing kink mode harmonics.

In order to study the heating of the plasma due to the phase-mixing of Alfv\'en waves, we first perform an analysis where
we fit the flux tube oscillation with the first three harmonics.
We additionally run a set of MHD simulations, where we vary the number of harmonics 
initially excited, as well at the width of the boundary shell.
Moreover, we also apply a seismology inversion to the 
MHD data in order to test the robustness of our model 
and to highlight what information can be derived 
from the seismology on the internal structure of coronal loops.

In the present study, we have found that 
the plasma heating that follows from the phase-mixing
of Alfv\'en waves is a marginal portion 
of the energy budget needed to maintain the thermal structure of coronal loops,
as the energy deposited is much smaller than the energy lost by radiation by the plasma,
where the model energy input was not arbitrary but constrained by the oscillations 
observed by \citet{2016A&A...589A.136P,2017A&A...600A..78P}.
Previously \citet{Cargill2016} and \citet{PaganoDeMoortel2017} raised concerns for the phase-mixing to be effective in providing the energy budget to support the coronal thermal structure.
At the same time, it is still possible that oscillations
are only a component of the process that leads to coronal heating which is ultimately enhanced by turbulence or reconnection phenomena, as suggested by several modelling studies, as \citet{Reale2016,Dahlburg2016}.
Therefore, it is still not possible to rule out the phase-mixing as heating mechanism without high spatial resolution and high cadence observations providing better estimates of the wave energy that is present in the solar corona \citep[e.g.][]{2012ApJ...746...31D}.

Nonetheless, some useful information can be gained from the analysis of the flux tube oscillation dynamics when higher order harmonics are included.
We find that the presence of multiple harmonics affects the heating of the flux tube in two ways.
Firstly, it affects the location of the heating the deposition.
The heating is mostly localised in the position along the flux tube
where more kinetic energy is initially available for conversion.
This is a consequence of the fact that the phase-mixing of Alfv\'en waves is a process that starts while the mode-coupling of the kink oscillation is still ongoing, i.e. the mode coupling and the phase-mixing are not occurring consecutively, but rather simultaneous.
Therefore, depending on the relative amplitude and phase of the harmonics involved, the heating is mostly localised in the position along the flux tube where the kink velocity profile peaks.
Moreover, as different harmonics have different damping times in the mode coupling process, the location of the 
velocity peak changes as higher harmonics disappear.
In our study this leads to a drift of the heating location towards the apex of the coronal loop, as the fundamental harmonic is the one with the longest damping time.

Secondly, when the kink oscillation kinetic energy is distributed among multiple harmonics, the heating process is less effective if KHI is also involved in the process.
In fact, we found that the coherent motion of one single harmonic leads to the highest velocity difference between the loop structure and the environment, thus allowing the strongest development of KHI in the boundary layer.
After the small scale structures are developed, the heating becomes more efficient.
In contrast, the interference between several harmonics is not always constructive and this results in lower oscillatory motion and thus slower formation of KHI.
A number of studies have already shown that KHI is sensitive to the relative motion between the oscillating structures and the environment 
\citep{BrowningPriest1984,Terradas2008,Foullon2011,Antolin2015,Okamoto2015,Magyar2016,Howson2017,Karampelas2017}.
Our work implies that the presence of higher order harmonics is one of the factors that inhibits or delays the formation of KHI and it should be taken into account when deriving loop properties from the development of small scale structures.
\citet{GoddardNakariakov2016} also found that smaller amplitude oscillations persist for longer, also suggesting that non-linear effects (such as the development of small scales) accelerate the damping of the oscillation.

\citet{Karampelas2017} and \citet{VanDoorsselaere2007a} argue that heating directly due to resistivity and viscosity can take place at different locations of the flux tube, as the first occurs through the dissipation of electric currents and the second of velocity shears.
Our simulation partially confirms this suggestion.
In fact, by comparing our simulation with a simulation where we set $\eta=0$ we find that the effect of resistivity is more visible at the foot-points (where currents induced by the oscillation are stronger)
than near the centre of the flux tube (where the current is smaller, but the velocity shear is higher).
At the same time, a known effect of mode-coupling of Alfv\'en waves is to eventually concentrate energy into a small region and the main effect of KHI is to further fragment these structures and to generate new currents , while the currents induced by oscillations become weaker because of the damping of the kink modes.
Therefore, it is reasonable to assume that in this scenario the heat deposition at centre of the flux tube eventually prevails on the footpoint heating.
This also suggests that our results are due to a combination of the effects of numerical viscosity whose impact increases with high velocity gradients and the effects of magnetic resistivity dissipating the small scale currents developed by phase-mixing and KHI.

It should also be noted that the line-tied boundary conditions used in MHD simulations to model the behaviour of loop footpoints probably lead to an overestimation of the electric currents generated at these locations by standing modes, as it has not been thoroughly investigated whether coronal loops are perfectly anchored at their coronal footpoints during kink oscillations.

It is also worth addressing the differences between our model and the observations which our MHD simulation parameters have been inspired by.
%While the initial oscillation amplitude, the period of oscillation, 
%the damping time of the 2nd and 3rd harmonics, and the values of $\epsilon$ and $R$ are correctly reproduced in our simulation
%and in its seismology inversion, the most notable difference is 
%the damping of the fundamental mode.
In our simulation the oscillation of the fundamental mode is still visible after $5P$, while it is not the case in the observation which is nearly completely damped after $4P$.
There are several modelling parameters that can justify this discrepancy that
cannot be resolved by the observations, but still affect the exact dynamics of the loop. One is the density profile we have assumed Sect.~\ref{sect:initial}.
%This is partly due to the different definition of the boundary shell width $\epsilon$ used in this study compared to the analytical method (see Sect.~\ref{sect:initial}).
However, seismological techniques developed thus far are also based on the thin boundary approximation, which is not applicable to our simulations specifically, and coronal loops in general \citep[e.g. statistical study by][]{2017A&A...605A..65G}.
The consequence of this approximation is an error of $\sim 20\%$ in current seismological inversions.
The damping rate in our simulations will also be affected by numerical effects due to finite resolution. %or by the
%spatial extension of the flux tube, as the seismology inversion 
%from observation is based on the thin boundary approximation whereas the MHD simulation considers spatially extended loop and boundary shell, or
%by a different density profiles, as the seismology uses a linear transition layer profile, whereas the MHD simulation uses a different one (Eq.~\ref{densitylayer}).
%\citet{2017A&A...600A..78P}

Finally, this study also hints that details of the loop internal structure, such as the radius of the internal structure or the width of the boundary shell, may be inferred from the seismological and forward modelling techniques, such as Fig.~\ref{ft} which shows changes in the estimates of $\epsilon$ and $R$ occur following the development of KHI.

\begin{acknowledgements}
This research
has received funding from the Science and Technology Facilities Council (UK)
through the consolidated grant ST/N000609/1 and the European Research Council
(ERC) under the European Union’s Horizon 2020 research and innovation
program (grant agreement No. 647214).
This work is supported by the European Research Council under the \textit{SeismoSun} Research Project No. 321141 (DJP).
This project has received funding from the European Research Council
(ERC) under the European Union's Horizon 2020 research and innovation
programme (grant agreement No 724326).
This work used the DiRAC Data Centric system at Durham University, operated by the Institute for Computational Cosmology on behalf of the STFC DiRAC HPC Facility (www.dirac.ac.uk. This equipment was funded by a BIS National E-infrastructure capital grant ST/K00042X/1, STFC capital grant ST/K00087X/1, DiRAC Operations grant ST/K003267/1 and Durham University. DiRAC is part of the National E-Infrastructure.
We acknowledge the use of the open source (gitorious.org/amrvac) MPI-AMRVAC software, relying on coding efforts from C. Xia, O. Porth, R. Keppens.
\end{acknowledgements}

%\begin{thebibliography}{}

\bibliographystyle{aa}
\bibliography{ref}

%\end{thebibliography}

\end{document}